\documentclass[a4paper,11pt]{article}

\pdfoutput=1
\usepackage{jheppub}

\usepackage{graphicx,color,amsmath}
\usepackage[utf8]{inputenc} 

\newcommand{\second}{{\, {\rm s}}}

\newcommand{\cm}{{\, {\rm cm}}}

\newcommand{\eV}{{\, {\rm eV}}}
\newcommand{\keV}{{\, {\rm keV}}}
\newcommand{\MeV}{{\, {\rm MeV}}}
\newcommand{\GeV}{{\, {\rm GeV}}}

\newcommand{\kelvin}{{\, {\rm K}}}
\newcommand{\gram}{{\, {\rm g}}}

\newcommand{\OO}{\mathcal{O}}

\newcommand{\erg}{{\, {\rm erg}}}

\newcommand{\egs}{\erg \gram^{-1} \sec^{-1}}
\newcommand{\gcmt}{\gram \cm^{-3}}

\newcommand{\Imag}{{\, {\rm Im}}}
\newcommand{\Real}{{\, {\rm Re}}}

\begin{document}

\title{Stellar cooling bounds on new light particles: plasma mixing effects}


\author[a]{Edward Hardy,}
\emailAdd{ehardy@ictp.it}
\affiliation[a]{Abdus Salam International Centre for Theoretical Physics,
Strada Costiera 11, 34151, Trieste, Italy}
\author[b]{Robert Lasenby}
\emailAdd{rlasenby@perimeterinstitute.ca}
\affiliation[b]{Perimeter Institute for Theoretical Physics, 31 Caroline Street N, Waterloo, Ontario N2L 2Y5, Canada}

\abstract{
	Strong constraints on the coupling of new light particles
	to the Standard Model (SM) arise from their production in the hot cores
	of stars, and the effects of this on stellar cooling. 
	For new light particles which have an effective in-medium mixing
	with the photon, plasma effects can result in parametrically different
	production rates to those obtained from a naive calculation.
	Taking these previously-neglected contributions into account, we make 
	updated estimates
	for the stellar cooling bounds on light scalars and vectors
	with a variety of SM couplings.
	In particular, we improve the bounds on light ($m \lesssim \keV$)
 	scalars coupling to electrons or nucleons by up to 3 orders
 	of magnitude in the coupling squared, significantly revise the
 	supernova cooling bounds on dark photon couplings,
	and qualitatively change the mass dependence of stellar bounds
	on new vectors. Scalars with mass $\lesssim 2 \keV$ that couple through the Higgs portal are constrained to mixing angle $\sin\theta \lesssim 3 \times 10^{-10}$, which gives the dominant bound for scalar masses above $\sim 0.2 \eV$.
}

\maketitle


\section{Introduction}

Many Beyond-Standard-Model (BSM) physics scenarios include new light, weakly-coupled
particles. If a new particle has mass $\lesssim 100 \MeV$, then it will
be produced in the hot cores of stars/supernovae, and will contribute
to energy transport in the star. The non-observation of anomalous energy transport,
in various different types of star, can then place strong constraints
on the coupling of the particle to the Standard Model (SM) particles
that make up the star~\cite{Raffelt:1996wa}.
These bounds are often referred to as `stellar cooling', though
they more properly correspond to anomalous energy loss or transport.

Previous calculations of particle production in stars
have often assumed that a simple `kinetic theory' calculation,
using in-vacuum matrix elements for processes producing the new state,
with thermal abundances for the SM initial and final states,
is a good approximation (\cite{Carlson:1986cu,Grifols:1988fv,Dent:2012mx,Rrapaj:2015wgs} are some examples). However, the large electron density
in stellar cores results in large characteristic plasma frequencies,
giving non-negligible collective effects.
These are especially important for new light particles that
gain an in-medium mixing with SM excitations.
In particular, if the new particle
is light compared to the thermal mass of the SM excitation,
then such mixing effects can parametrically change the production
rate, compared to the naive kinetic theory estimate.

These phenomena have been taken into account for dark
photon emission from stars (not including supernovae)
in~\cite{An:2013yfc,Redondo:2013lna}, after their possible importance was first identified by~\cite{Redondo:2008ec}.
In this work, we point about that similar effects hold for other
forms of BSM particle interactions, and can lead to important
changes in stellar cooling constraints. 
In the case of dark photon, these effects can be included
by choosing a sterile/active basis, in which the sterile mode
couples to the medium purely through its Lagrangian mass mixing.
However, such a basis choice is not generically possible, and
we explain how to compute production rates in more general cases.
We also extend the analysis of dark photon production
to include energy loss from core-collapse supernovae,
correcting previous literature which used the kinetic theory
approximation~\cite{Rrapaj:2015wgs,Kazanas:2014mca,Dent:2012mx}, and to take into
account trapping constraints in stars. The resulting constraints are shown in Figure~\ref{fig:sn1}.

New light vectors which couple coherently to SM plasma oscillations have an in-medium mixing with the SM photon. At
vector masses well below the stellar plasma frequency, this mixing 
suppresses the production rate. At masses around the plasma frequency,
resonant conversion of SM photons to the new vector can enhance the
production rate, while at masses above the plasma frequency the kinetic
theory approximation holds. Figure~\ref{fig:bl1} shows how this changes
the solar bounds on a massive $B-L$ vector.

New light scalars can mix with SM plasmons --- the in-medium
`longitudinal mode of the photon'~\cite{Raffelt:1996wa}. Since the plasmon dispersion
relation crosses the light-cone, resonant production of the light scalar
is possible down
to arbitrarily low scalar masses, and unlike the case for a light
vector, is not suppressed by powers of the mass.
So, for scalar masses below the stellar plasma frequency, 
resonant production can dominate, parametrically strengthening
the stellar cooling bounds.
Figure~\ref{fig:hp1} shows
this effect for a light Higgs-portal scalar.
Quantitatively, we improve previous coupling bounds by a
factor of $\sim 10^3$ (in the coupling squared) for both a $\phi \bar{e} e$ coupling,
and a scalar coupling to nucleons. 

While
light scalar or vectors are the simplest cases,
mixing effects may also be important for other new particle candidates,
the detailed investigation of which we leave to future work.
For some candidates, symmetries
of the low-energy theory prevent an in-medium mixing
with SM states (e.g.\ charge conservation for a milli-charged
particle, or parity for an axion-like particle). However,
plasma effects may still have an important impact
on thermal production rates of such particles
(via screening, thermal masses, etc.), although we will not discuss
them here.
In addition to stellar cooling, there are also other physical scenarios
in which our calculations may be useful, among them
early-universe cosmology.

The main aim of this paper is to illustrate how plasma mixing effects
may have important consequences for the thermal production
of new light particles, not to calculate precise and robust
constraints on such particles. Consequently, we do not
do the detailed stellar modelling that would be required
to derive properly reliable bounds, instead taking representative
stellar models and approximate analytic versions of observational
bounds from the literature. However, this suffices to demonstrate
the parametric changes that our new physical effects
bring about, and shows that they would have important
effects in a detailed analysis.

The structure of this paper is as follows: In Section~\ref{sec:thermalprod} we
introduce the machinery of thermal field theory,
and obtain the proper expressions for production rates of new light BSM
states. Following this, in Section~\ref{sec:bounds} we apply these results
to stellar cooling, and update the constraints on BSM couplings from
observations. Finally, in Section~\ref{sec:discuss},
we discuss our results, and review directions for future work.


\section{Thermal production rates and plasma mixing}
\label{sec:thermalprod}

The conceptually-simplest way to handle medium effects in a thermalised
plasma, such as that found in a stellar core, is to use the apparatus of thermal
field theory (TFT) --- for a comprehensive overview, see e.g.~\cite{Bellac:2011kqa}.
In this framework, 
weakly-coupled excitations in a medium correspond to poles
in the thermal propagators of fields, with the decay and production
rates of these excitations being related to the imaginary parts of the pole locations.
If we have a state $X$ that couples weakly to the species making up
a thermalised bath, then writing the thermal propagator
in terms of the free propagator and the `thermal self-energy',
$D_X = \frac{1}{D_X^F - \Pi_X}$, we are interested in the imaginary
part of $\Pi_X$
(see Appendix~\ref{ap:tft} for a short review of propagators in the real-time thermal
field theory formalism).

\begin{figure}
	\begin{center}
		\includegraphics[width=.8\textwidth]{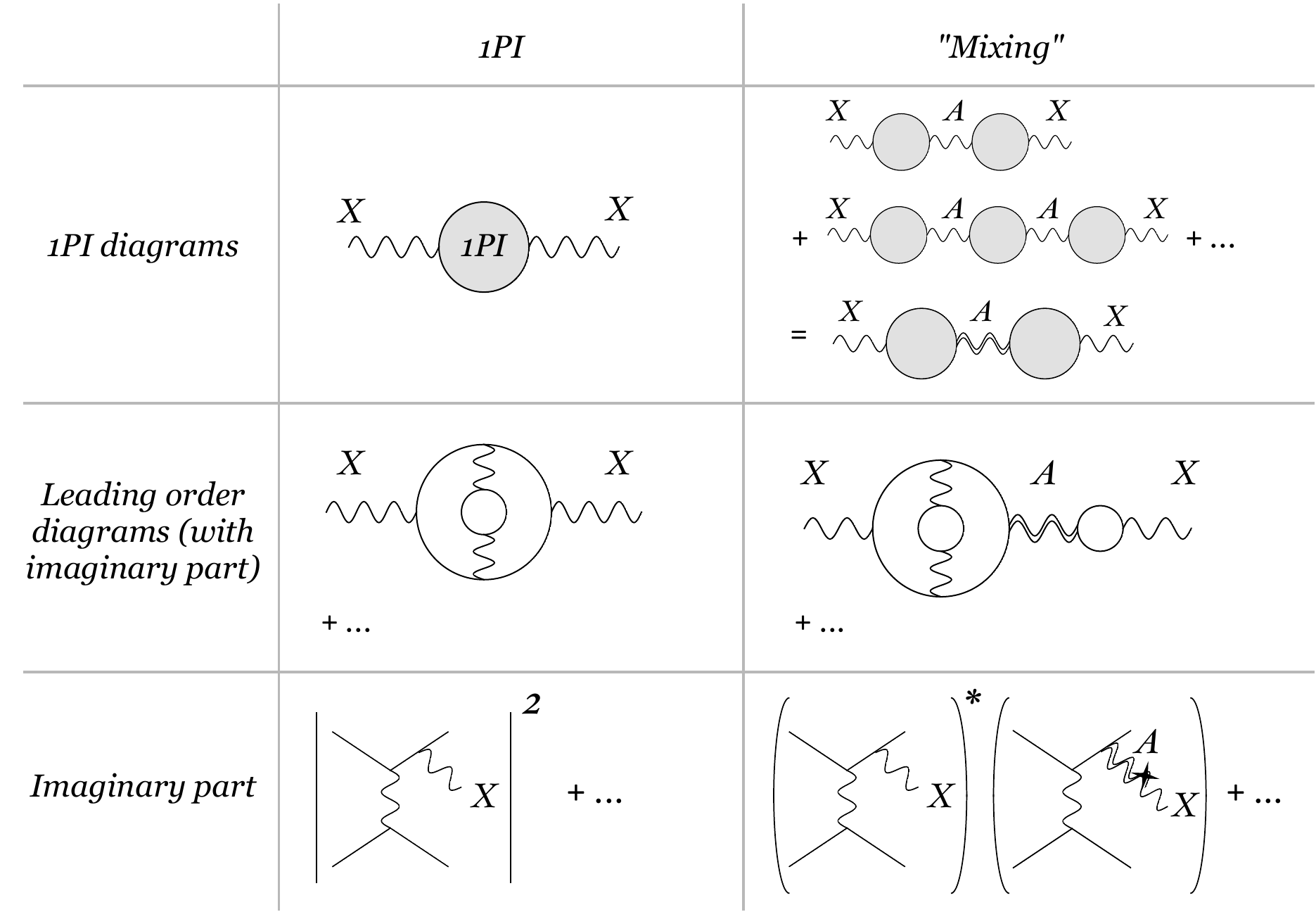}
	\end{center}
	\caption{
		Illustration of the 1PI contributions to the thermal field theory
		self-energy of a particle $X$, versus
		the `mixing' contributions involving intermediate
		bath species propagators, where we take $X$ to mix with
		a single bath species $A$.
		The double-lined $A$ propagator indicates the full (resummed) TFT
		propagator.
		The third row shows some of the tree-level
		contributions to the imaginary part of the self-energy,
		with the `mixing' diagrams corresponding to $X$
		production involving its medium-induced mixing with $A$
		(indicated by the star).
		Many calculations of thermal particle production in the literature 
		only consider the class
		of diagrams show in the 1PI column,
		ignoring the `mixing' contributions.
		}
	\label{fig:mixing1}
\end{figure}

This paper's central point is that, if $X$ mixes with the bath fields, either through
terms in the Lagrangian, or through medium-induced effects, then $\Pi_X$
is not just given by the sum of one-particle-irreducible (1PI) TFT diagrams.
Instead, diagrams with intermediate bath species propagators must also be
included. This is illustrated, in the case where $X$ mixes with a single
bath species $A$, by Figure~\ref{fig:mixing1}.
At the level of a leading-order `kinetic theory' calculation, the extra
terms correspond to including $X$ production through its medium-induced mixing
with $A$.
This point may seem trivial, but we will see that many calculations
in the literature correspond to ignoring mixing effects,
which can lead to parametrically incorrect results.

To see the effects of mixing more explicitly, we can consider the
species-non-diagonal terms. Supposing again that $X$ mixes with a single bath
species $A$ (in the models we consider, $A$ will generally be the SM photon), 
the in-medium propagation eigenstates
will be given by null eigenvectors of the $2 \times 2$ matrix
\begin{equation}
	\begin{pmatrix}
		K^2 - \Pi^{AA} & -\Pi^{AX} \\
		-\Pi^{XA} & K^2 - m_X^2 - \Pi^{XX}
	\end{pmatrix} ~,
\end{equation}
where $K = (\omega, \vec{k})$ is the 4-momentum,
and we have taken $m_A = 0$. Here, the self-energies
$\Pi^{AX}$ etc.\ \emph{do} correspond to the sum of 1PI diagrams
(evaluated at real energy-momentum).
If the couplings of $X$
to bath states are $\OO(g)$, where $g$ is small, then since
$\Pi^{AX} = \OO(g)$ and $\Pi^{XX} = \OO(g^2)$,
for a given $\vec{k}$ there exist null eigenvectors when
\begin{equation}
	\omega_c^2 = k^2 + \Pi^{AA} + \frac{(\Pi^{AX})^2}{\Pi^{AA} -m^2} + \OO(g^4)~,
\end{equation}
\begin{equation}
	\omega_c^2 = k^2 + m^2 + \left( \Pi^{XX} - \frac{(\Pi^{AX})^2}{\Pi^{AA} - m^2}
	\right) + \OO(g^4) ~,
\end{equation}
where $\omega_c \equiv \omega + i \omega_i$ is the complex frequency.
The second expression corresponds to the weakly-coupled state, and
we can see how the self-energy corresponds to the sum of the terms
from Figure~\ref{fig:mixing1}.
These expressions make sense if $g$ is small compared to all other parameters.
In particular, if $\Pi^{AA} - m^2$ becomes very small (`on-resonance'),
they may cease to apply, as discussed in Section~\ref{sec:dp}.
The canonically-normalised propagation eigenstates are, in the $(A,X)$ basis,
\begin{equation}
	\sqrt{Z_A^{-1}}\begin{pmatrix} 1 \\ \frac{\Pi^{AX}}{\Pi^{AA} - m^2} 
	\end{pmatrix}+ \OO(g^2)
	\quad , \quad
	\begin{pmatrix} - \frac{\Pi^{AX}}{\Pi^{AA} - m^2} \\ 1
	\end{pmatrix}+ \OO(g^2)~,
\end{equation}
where
\begin{equation}
	Z_A^{-1} \equiv 1 - \frac{d \Real \Pi}{d \omega^2} ~,
\end{equation}
is the wavefunction renormalisation factor for $A$,
with the derivative being evaluated on-mass-shell.
The renormalisation
factor for the mostly-$X$ eigenstate is $1 + \OO(g^2)$.

The mostly-$X$ eigenstate has at least $g$-suppressed
interactions with the thermal bath. So, if the bath is of finite
size (such as a star) and $g$ is small enough, it will free-stream out of
the bath, rather than coming to thermal equilibrium.
In this case, we are interested in the production
rate of $X$ by the bath. If the order-$g$ couplings
of $X$ to the bath involve production/absorption of one $X$,
rather than scattering of $X$, then the production rate can be related,
via detailed balance, to the damping
rate given by the imaginary part of the frequency.
Since the production and absorption rates satisfy
$\Gamma_{\rm prod} = e^{-\omega/T} \Gamma_{\rm abs}$,
and the overall damping rate is $\Gamma = \Gamma_{\rm abs} - \Gamma_{\rm prod}$
(for a bosonic excitation),
we have $\Gamma_{\rm prod} = f_B(\omega) \Gamma$, where
$f_B(E) \equiv (e^{E/T} - 1)^{-1}$ is the bosonic thermal
occupation number.
The production
rate per volume for the (assumed bosonic) mostly-$X$ state is therefore given by 
\begin{equation}
	\frac{dN_{\rm prod}}{dV dt} = \int \frac{d^3 k}{(2\pi)^3} \Gamma_{\rm prod}
	= - \int \frac{d^3 k}{(2 \pi)^3} \frac{f_B(\omega)}{\omega}
	\Imag \left( \Pi^{XX} - \frac{(\Pi^{AX})^2}{\Pi^{AA} - m^2}
	\right) + \OO(g^4) ~.
	\label{eq:prodrate1}
\end{equation}
The essential point of this paper is that, for production of new
weakly-interacting particles from a SM bath, many previous analyses
correspond to only considering the $\Pi^{XX}$ term, whereas in fact
the other term can sometimes be very important.\footnote{
	An additional point is that, though we have calculated
	these `mixing effects' within the thermal field theory framework,
	they are better thought of as `plasma' effects rather than `thermal ones',
	and will also occur out of thermal equilibrium. Appendix~\ref{ap:eom}
works out a particular case by considering
	classical plasma oscillations in the fluid approximation,
	with the presence of a new weakly-coupled field in addition to electromagnetism;
	this gives a toy example of a non-thermal calculation of this kind.
	}
One straightforward example where this cannot be ignored is
the case of low-mass dark photon production, as has been noted in a number
of papers~\cite{Redondo:2008ec,An:2013yfc,Redondo:2013lna}. We will see that this also occurs in other
cases of phenomenological interest.


\subsection{SM plasma oscillations}
\label{sec:smplasma}

Suppose that there exists an as-yet-undiscovered light bosonic state,
which couples weakly to the SM (a new fermionic state, if it is not
detectably SM-charged, can only mix with the SM neutrinos, and such a
mixing would not be important in most scenarios we consider). If this
state does not carry SM quantum numbers,
then, in the TFT treatment of a low-energy ($T \ll \Lambda_{\rm QCD}$)
SM plasma, it can only mix with the photon field.
In this section, we review the photon field's TFT behaviour (more comprehensive reviews
can be found in~\cite{Raffelt:1996wa,Braaten:1993jw}).

\begin{figure}
	\begin{center}
		\includegraphics[width=.48\textwidth]{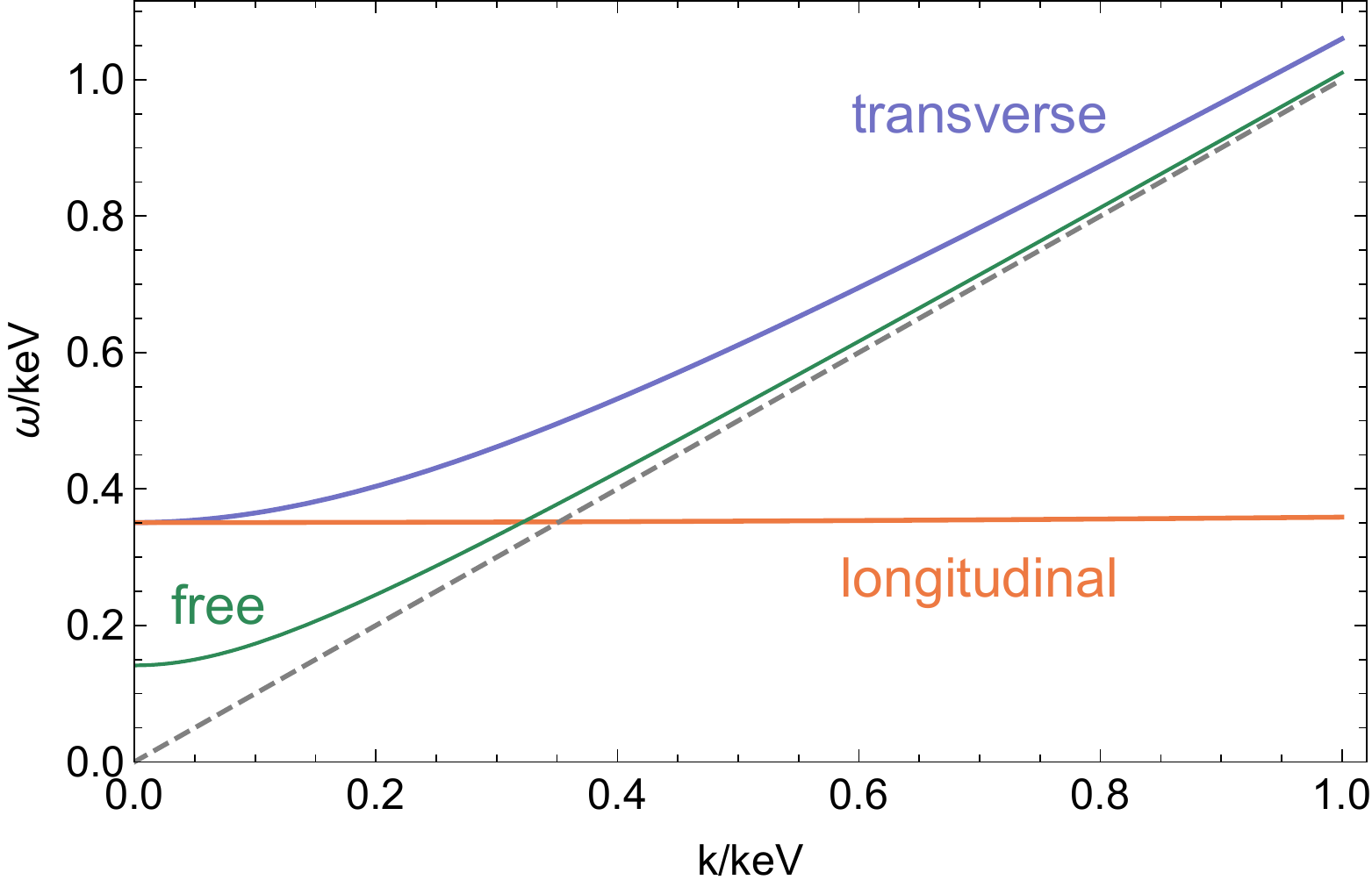}
		\includegraphics[width=.48\textwidth]{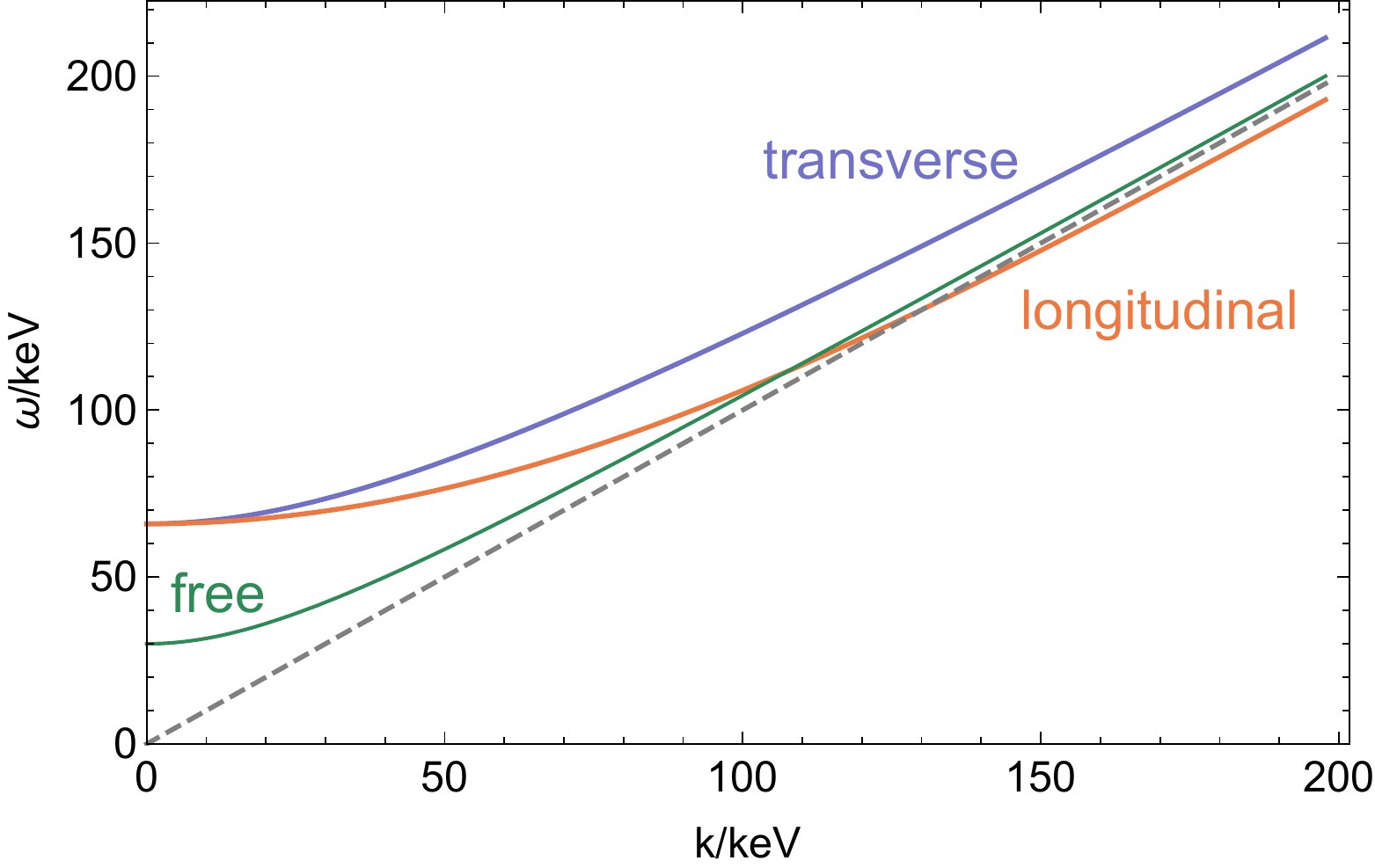}
	\end{center}
	\caption{
		The photon dispersion relation in solar core plasma (left)
		and early universe plasma at $T = 1 \MeV$ (right),
		showing non-relativistic and relativistic dispersion
		relations respectively.
		Blue lines show transverse dispersion relations,
		orange lines show longitudinal.
		For both plasmas, transverse and longitudinal oscillations
		are weakly damped, resulting in narrow resonances
		around the plotted frequencies.
		The grey dotted lines indicate the light cone.
		Green lines show example dispersion relations for weakly-coupled
		light particles ($m = 140 \eV$ on left, $m = 30 \keV$ on right).
		Since, as discussed in Appendix~\ref{ap:photonpi},
		the effects of increasing temperature and chemical potential
		once $T,\mu \gg m_e$ are almost degenerate,
		the right-hand plot is qualitatively similar to the shape
		of the dispersion relations in e.g.\ a SN core.
		}
	\label{fig:disp1}
\end{figure}

We can write the in-medium photon self-energy, in an isotropic medium, as
\begin{equation}
	\Pi_{\mu\nu}(K) = (\epsilon^{+}_\mu {\epsilon^{+}_\nu}^*
	+ \epsilon^{-}_\mu {\epsilon^-_\nu}^*)
	\Pi_T + \epsilon^L_\mu
	\epsilon^L_\nu \Pi_L ~,
\end{equation}
where $\epsilon^{\pm}$ are the usual orthogonal polarisation vectors~\cite{Raffelt:1996wa},
and $\epsilon^L_\mu = \frac{1}{\sqrt{K^2}} (-|k|, \omega \hat{\vec{k}})$
is the longitudinal polarisation vector (working in the rest frame of the medium).
We will write $\Pi = \Pi_r + i \Pi_i$ for the real and imaginary parts
of the self-energy.
For a non-relativistic, dilute plasma, the real parts are~\cite{Raffelt:1996wa}
\begin{equation}
	\Pi_{T,r} = \omega_p^2 \left(1 + \frac{k^2}{\omega^2} \frac{T}{m_e}\right)   \quad , \quad
	\Pi_{L,r} = \frac{K^2}{\omega^2}\omega_p^2 \left(1 + \frac{3 k^2}{\omega^2} \frac{T}{m_e}\right) ~,
\end{equation}
to first order in the small average electron velocities, 
where
\begin{equation}
	\omega_p^2 = \frac{e^2 n_e}{m_e}\left(1 - \frac{5}{2}\frac{T}{m_e} + 
	\OO\left(\frac{T^2}{m_e^2}\right)\right)
	+ \OO\left(\frac{m_e}{m_n}\right) ~,
\end{equation}
with $T$ the temperature of the medium and $n_e$ the electron number density.
Figure~\ref{fig:disp1} plots the corresponding dispersion relations.
The essential physics is that the usual transverse modes are simply
modified by the presence of a small effective plasma mass,
whereas the longitudinal modes oscillate at an almost-fixed
frequency set by the same plasma mass, propagating only through
thermal diffusion effects.

This unusual dispersion relation for the longitudinal modes has
the very important consequence that, for a light
new particle of mass $m < \omega_p$, there is always a $k$ where the
dispersion relations of the new particle and the longitudinal photon cross.
This means that the denominator in equation~\ref{eq:prodrate1} can become small,
allowing resonant production of the new state,
as pointed out in the case of a dark photon by~\cite{Redondo:2008aa}.
Resonant production from mixing with transverse photons is also possible,
but only for $m$ in a narrow range around $\omega_p$. 

The situation changes somewhat for relativistic or degenerate plasmas.
As we increase the typical electron velocity, the propagation speed of
the longitudinal mode increases, moving the point where the longitudinal
dispersion relation crosses the lightcone to higher $k$. In the
ultra-relativistic limit, both dispersion relations are always above
the lightcone. However, the ratio of the cross-over point to the plasma
frequency increases only logarithmically with the typical electron
energy, so even for the highly relativistic plasma in a supernova core,
cross-over still occurs at $\omega$ only a few times the temperature.
Thus, there is still the possibility of resonant production for very
light, weakly-coupled states through mixing with the longitudinal mode.
Details are given in Appendix~\ref{ap:photonpi}.

One point to note is, well below the light-cone, the photon self-energy
will have more complicated behaviour, and the longitudinal propagator will
generally have extra poles corresponding to ion acoustic oscillations etc.\
(c.f.~\cite{Chen1984}). However, we are interested in the emission of
weakly-coupled massive particles, so in the behaviour above the light-cone
--- in particular, resonant production can only occur for the
longitudinal and transverse photon oscillations.

For notational convenience, we define $m_T(K)^2 \equiv \Pi_{T,r}(K)$,
$(K^2/\omega^2)\omega_L(K)^2 \equiv \Pi_{L,r}$; 
the longitudinal definition factors out the $K^2/\omega^2$
dependence that appears automatically from the polarisation contractions.
From the previous section, the width for transverse
modes is $\Gamma_T(K) \simeq -\Pi_{T,i} / \omega$ in
the dilute non-relativistic limit
(in fact, as per~\cite{Braaten:1993jw},
this is always true to good accuracy), and for longitudinal
modes we write $-\Pi_{L,i} \equiv (K^2/\omega^2) \omega \sigma_L(K)$,
where this equation defines $\sigma_L(K)$.
In the non-relativistic case, $Z_L^{-1} = 1 - \frac{d \Pi_r}{d\omega^2} \simeq K^2/\omega_p^2$,
so, the physical width of longitudinal plasmons
is $\Gamma_L \simeq (Z_L/\omega_p) (-\Pi_{L,i}) = \sigma_L$.


\subsection{Dark photons}
\label{sec:dp}

Dark photons provide the simplest example of a new state
mixing with SM photons. The physics in this case has been
described in previous literature~\cite{An:2013yfc,Redondo:2013lna},
but we go over it here to show the similarity with other cases to follow,
and since we will extend the dark photon constraints to include
SN cooling and stellar trapping. 

Suppose we have a new vector $A'$, the `dark photon', which couples
to the SM EM current $J$,
\begin{equation}
	\mathcal{L} \supset
	- \frac{1}{4}F^2 - \frac{1}{4}F'^2 + \frac{1}{2}m^2 A'^2 
	+ e J (A + \epsilon A') ~.
\end{equation}
Note that, after a field redefinition,
this is equivalent to kinetic mixing 
$
	\mathcal{L} \supset
	- \frac{1}{4}F^2 - \frac{1}{4}F'^2 - \frac{\epsilon}{2} F F' + \frac{1}{2}m^2 A'^2 
	+ e J A
	$.
We will assume that there are no additional BSM states
at low energies, so that the mass $m$ is a Stueckelberg
mass (this is natural even with very small $m$, since $J$ is a conserved
current), rather than coming from a low-scale Higgs mechanism.
Then, $\Pi^{XX} = \epsilon^2 \Pi^{AA}$, $\Pi^{AX} = \epsilon \Pi^{AA}$, so
writing $\Pi^{AA} \equiv \Pi = \Pi_r + i \Pi_i$,
\begin{equation}
	- \Imag(\omega_c^2) = \Imag \left( \epsilon^2 m^2 \frac{\Pi}{\Pi - m^2}\right) + \OO(\epsilon^4)
	= \epsilon^2 \frac{m^4 (-\Pi_i)}{\Pi_i^2 + (\Pi_r - m^2)^2}
	+ \OO(\epsilon^4) ~.
\end{equation}
In~\cite{An:2013yfc,Redondo:2013lna}, this result was obtained
by going to the active/sterile basis, in which the sterile
state couples to the SM only via its mass mixing with the active state.
As per Section~\ref{sec:thermalprod}, we then have no $\Pi^{XX}$ 1PI contribution,
but just the $(\Pi^{AX})^2/(\Pi^{AA} - m^2)$ term.
The end result is, of course, independent of the basis chosen, as illustrated
by our choice of the (vacuum) mass basis. 

If the SM photon oscillations are weakly damped, $\Pi_i \ll \Pi_r$
(as is generally the case for the environments of interest here),
we can split the production rate into four different contributions, with parametrically
different properties:

\begin{itemize}

	\item Continuum transverse production: the production of transverse weakly-coupled
		states has rate
		\begin{equation}
			\Gamma = \frac{1}{e^{\omega/T}-1} \epsilon^2 m^4 \frac{\Gamma_T}{\omega^2 \Gamma_T^2 + (m_T^2 - m^2)^2} ~.
		\end{equation}
		For $|m_T^2 - m^2| \gg \omega \Gamma_T$, i.e.\ off-resonance,
		this is
		\begin{equation}
			\Gamma \simeq \epsilon^2 \frac{m^4}{(m_T^2 - m^2)^2} \Gamma_{T,{\rm prod}}~.
		\end{equation}
		For $m \gg m_T$, this is simply the `naive' production
		rate $\epsilon^2 \Gamma_{T,{\rm prod}}$, while for $m \ll m_T$, it is suppressed
		by $m^4/m_T^4$. In an active/sterile basis, this suppression
		comes from the propagator for the heavy active state~\cite{An:2013yfc}.

	\item Resonant transverse production: for $|m_T^2 - m^2| < \omega \Gamma_T$,
		the production rate is~\footnote{
			As mentioned in Section~\ref{sec:thermalprod}, this expression holds
			if $\epsilon$ is small compared to $(\Pi - m^2)/\Pi$.
			In particular, at $\Pi_r = m^2$, if $\epsilon > \frac{1}{2} \omega \Gamma_T / m^2$,
			then to leading order in $\omega \Gamma_T/m^2$, we have
			$\Gamma \simeq (1/2)\Gamma_{T, {\rm prod}}$,
			so the dark photon production rate never parametrically exceeds 
			that for the SM photon.
			}
		\begin{equation}
			\Gamma \simeq  \frac{1}{e^{\omega/T}-1}  \epsilon^2 \frac{m^4}{\omega^2 \Gamma_T} ~.
		\end{equation}
		If we consider production at a given frequency $\omega$,
		then for a spatially-varying medium, there is some region
		for which the resonance condition holds.
		The size of this region will be set by $\sim \omega \Gamma_T /(\partial_x m_T^2)$,
		so the overall power emitted will be \emph{independent}
		of $\Gamma_T$, in the small-width approximation.
		If we assume a spherically-symmetric medium, with
		$m_T$ a function of $\omega,k,r$, we have~\cite{Redondo:2008aa}
		\begin{equation}
			P
			= \int dr\, 4\pi r^2 \int \frac{d^3 k}{(2 \pi)^3} 
			\frac{1}{e^{\omega/T} - 1} \epsilon^2 m^4 \frac{(-\Pi_i)}{(\Pi_i)^2
			+ (m_T^2 - m^2)^2}
		\end{equation}
		\begin{equation}
			\simeq \int d\omega\, 2 k \omega
			\frac{1}{e^{\omega/T} - 1} \epsilon^2 m^4 \left(r^2
			\left|\frac{\partial m_T^2}{\partial r}\right|^{-1}
			\right)_{r \, {\rm s.t.} \, m_T^2(\omega,k,r) = m^2} ~,
		\end{equation}
		where the integral is
		over $\omega$ such that
		we match the resonance condition at some $r$. 

	\item Continuum longitudinal production: the production of longitudinal
		weakly-coupled states has rate
		\begin{equation}
			\Gamma = \frac{1}{e^{\omega/T}-1} \epsilon^2 m^2 \omega^2 \frac{\sigma_L}{\omega^2 \sigma_L^2 + (\omega^2 - \omega_L^2)^2} ~.
		\end{equation}
		Off-resonance, this is 
		\begin{equation}
			\Gamma \simeq \frac{1}{e^{\omega/T}-1} \epsilon^2 \frac{m^2 \omega^2}{(\omega^2 - \omega_L^2)^2} \sigma_L ~.
		\end{equation}
		As expected, the production rate is always suppressed by $m^2$
		(since $\sigma_L$ has already had the leading $m^2$ dependence factored
		out). Since the EM current is conserved, longitudinal emission
		is always suppressed by $m^2$, in vacuum and in medium~\cite{An:2013yfc}.

	\item Resonant longitudinal production: for $|\omega^2 - \omega_L^2| < \omega \sigma_L$,~\footnote{
			Analogously to transverse resonant production, this expression is valid
			for $\epsilon \ll \sigma_L/\omega$. If $\epsilon > (1/2)\sigma_L/\omega$,
			then to leading order in $\sigma_L/\omega$, 
			we have $\Gamma \simeq (1/2) (e^{\omega/T}-1)^{-1} (m^2/\omega^2) \sigma_L$,
			so the resonant longitudinal production rate is always down
			by a factor of $m^2/\omega^2$ relative to $\sigma_L$.
			}
		\begin{equation}
			\Gamma \simeq  \frac{1}{e^{\omega/T}-1}  \epsilon^2 \frac{m^2}{\sigma_L} ~.
		\end{equation}
		As noted in Section~\ref{sec:smplasma}, this differs
		from the resonant transverse case in that, if $m < \omega_L$,
		there is always a $k$ such that the resonance condition holds.
		Therefore, we can find the emissivity at a given position,
		at leading order in the small-$\sigma_L$ approximation,
		via
		\begin{equation}
			\frac{d\dot N_{\rm prod}}{dV} = \int \frac{d^3 k}{(2\pi)^3} \Gamma_{\rm prod}
			= \int \frac{2 k \omega d\omega}{(2 \pi)^2} \frac{1}{e^{\omega/T}-1}
			\epsilon^2 m^2 \omega \frac{\omega \sigma_L}{(\omega \sigma_L)^2 + (\omega^2 - \omega_L^2)^2}
		\end{equation}
		\begin{equation}
			\simeq
			\frac{1}{4\pi} \epsilon^2 m^2 k_{\omega_L} \omega_L \frac{1}{e^{\omega_L/T} - 1}
			\left| 1- \left. \frac{d\omega_L^2}{d\omega^2}\right|_{\omega_L}\right|^{-1} ~.
			\label{eq:resL}
		\end{equation}

\end{itemize}

As expected, at large masses $m \gg m_T, \omega_L$ we are in the kinetic theory regime
for longitudinal and transverse emission.
At masses small compared to the plasma frequencies, longitudinal production is
dominant, being suppressed by $m^2$.

\subsection{$B-L$ vector}
\label{sec:blprod}

As well as the electromagnetic current, there is another
non-anomalous (assuming Dirac neutrinos) conserved current
in the SM corresponding to $B-L$. A light vector coupling weakly to this
current,
\begin{equation}
	\mathcal{L} \supset
	-\frac{1}{4} X_{\mu\nu}X^{\mu\nu} + \frac{1}{2} m^2 X^2 + g J_{B-L} X ~,
\end{equation}
where $J_B$ is the SM $B-L$ current,
is a popular candidate
for new vector portal physics, e.g.~\cite{Alekhin:2015byh}.
Since $B-L$ does not have any mixed anomalies with the SM gauge groups,
there are no order-$g^2$ production processes that are enhanced
by powers of $E^2/m^2$, where $E$ is the energy of the process.
This means that a $B-L$ vector avoids the tight constraints,
coming from high-energy experiments,
that apply to vectors coupling to non-conserved currents.~\footnote{If
SM neutrinos are Majorana, then there is a $(B-L)^3$ anomaly ---
however, the cutoff scale implied by this anomaly is extremely high, and
it is not a phenomenological concern.}
In this paper, we will suppose that the only light
fermions coupling to $X$ are those of the SM
(the phenomenology we study is not significantly affected by whether
neutrinos are Dirac or Majorana), and that $X$ has a Stueckelberg mass.

In a medium consisting entirely of protons and electrons,
the tree-level behaviour of a $B-L$ vector is exactly the same
as for a dark photon. We therefore expect differences
in thermal production of a light $B-L$ vector vs a dark photon 
to be driven by the neutron content of the medium.

For a dilute, non-relativistic plasma, the dominant contribution
to the real parts of the $X$ and $A$ self-energies is
from the electrons, and the imaginary parts are also dominated
by emission from electrons (either electron-ion bremsstrahlung, or Compton
scattering). So, we can write 
\begin{equation}
	\Pi^{XX}_{r,i} = \frac{g^2}{e^2} \Pi^{AA}_{r,i} (1 + \alpha_{r,i}) ~,
\end{equation}
\begin{equation}
\Pi^{AX}_{r,i} = \frac{g}{e} \Pi^{AA}_{r,i} (1 + \beta_{r,i}) ~,
\end{equation}
where $\alpha$ and $\beta$ are suppressed by some function of the small
parameter $m_e/m_i$. Then,
\begin{align}
	-	\Imag (\omega_c^2) = 
	\frac{g^2 / e^2}{\Pi_i^2 + (\Pi_r - m^2)^2} (-\Pi_i)
		& \Big( 
 	m^4 (1 + \alpha_i) 
	+ 2 m^2 (\beta_r + \beta_i - \alpha_i) \Pi_r \nonumber  \\
	&  + (\Pi_r^2 + \Pi_i^2) (\alpha_i - 2 \beta_i)
	\Big) ~,
\end{align}
where $\Pi \equiv \Pi^{AA}$ is the photon self-energy, and we have ignored higher order terms
in $\alpha, \beta$.
As expected, the only part that is not $\alpha,\beta$
suppressed is the dark photon rate $m^4 \Pi_i$, which will dominate
at high masses, when the $m^4$ factor does not suppress it.
In particular, we have the same resonant production contributions as
for a dark photon, which were not visible in the `kinetic theory'
rate corresponding to $\Pi_i^{XX} = (g^2/e^2)\Pi_i^{AA} (1 + \alpha_i)$.

The part of the production that is unsuppressed as $m \rightarrow 0$ depends
on $\alpha_i - 2 \beta_i + \OO(\alpha^2, \beta^2)$
(for transverse emission; longitudinal emission is always suppressed by $m^2$).
The $\alpha_i$ and $\beta_i$ contribution from electron-ion bremsstrahlung
are $\OO(m_e/m_n)$ (arising from interference between
emission from electrons and from ions), but these cancel
at leading order, $\alpha_i - 2 \beta_i = \OO(m_e^2 / m_n^2)$.
This becomes clear if we go to a basis $\tilde{A} = A + (g/e) X$,
$\tilde{X} = X - (g/e) A$, where $\tilde{X}$ only couples to neutrons
--- then, 
\begin{equation}
	\Pi^{\tilde{X} \tilde{X}}_i = \Pi^{XX}_i - 2 (g/e) \Pi^{AX}_i + (g/e)^2 \Pi^{AA}_i
	= (g/e)^2 \Pi^{AA}_i (\alpha_i - 2 \beta_i) ~,
\end{equation}
and $\tilde{X}$ production in electron-ion bremsstrahlung is suppressed
by $m_e^2 / m_n^2$, since $\tilde{X}$ can only be emitted from the ion leg.

Compton production has no interference between emission from electrons 
and ions, and has $\alpha_i, \beta_i = \OO(m_e^2/m_n^2)$ at leading order.
The lowest-order in $m_e/m_n$ contribution to $m$-independent
production therefore comes from bremsstrahlung between different ion species,
which is suppressed by $(m_e/m_n)^{3/2}$.
At masses well below the plasma frequencies, this contribution
will dominate the production, while at masses comparable or above,
the dark-photon-like contribution is largest.~\footnote{
	\label{foot:blprod}
As discussed in Appendix~\ref{ap:damp}, for sufficiently low-temperature
plasmas, there is the additional complication that ion-ion bremsstrahlung
is further suppressed by the ratio $v_i/\alpha$, where $v_i \sim \sqrt{T/m_i}$
is the typical velocity of an ion, if this ratio is $\ll 1$.
This occurs since the Coulomb interaction between ions is repulsive,
and for low velocities, prevents them from getting within a de Broglie
wavelength of each other, so limiting
the accelerations they feel in the collision. For
example, in the solar core, the velocity of protons is $\sim 10^{-3} \ll \alpha$.
In this case, the ion-ion bremsstrahlung contribution is 
suppressed by $\sqrt{T} m_e^{3/2} / (m_n^2 \alpha)$.
Since $\sqrt{m_e/m_n} \sim 0.02$, there is the potential that
$m_e^2/m_n^2$ contributions may be numerically important in such circumstances.
For sufficiently dense plasmas, plasma screening
will also be important, since the relevant ion separations during collisions
is large due to Coulomb repulsion.
}

The production of $B-L$ bosons in relativistic and/or degenerate
plasmas is more complicated, and we leave it to future work.


\subsection{$\phi \bar{f} f$ scalar}
\label{sec:scalarprod}

A scalar with renormalisable couplings to SM fermions,
\begin{equation}
	\mathcal{L} \supset \frac{1}{2} (\partial_\mu \phi)^2 - \frac{1}{2} m^2 \phi^2
	+ \sum_f g_{\phi f} \phi \bar{f} f ~,
\end{equation}
in the low-energy theory, will also couple coherently to SM plasma
oscillations. This action can arise, for example,
from mixing with the SM Higgs.

The low-energy SM does not contain any scalar states, so
vacuum $\phi$-SM mixing is not possible. However, in a plasma,
$\phi$ can mix with the in-medium `longitudinal photon' mode.

For non-relativistic (relative to $f$) momentum transfers, 
$\bar{f}f \simeq \bar{f} \gamma^0 f$, so the contribution 
to the mixing self-energy from $f$ will be
\begin{equation}
	\Pi^{\phi A}_\mu \simeq \frac{g}{e Q_f} \Pi^{AA}_{0\mu}
	\quad \Rightarrow \quad
	\Pi^{\phi A}_\mu (\epsilon^L)^\mu  \simeq \frac{g}{e Q_f}  \frac{k}{\sqrt{K^2}} \Pi_L ~,
\end{equation}
where $Q_f$ is the EM charge of $f$.
Thus, on mass-shell, there is a $\sim k/m$ enhancement over the longitudinal
vector self-energy. As we will see below, this will translate into a different
mass dependence of the overall production rate.

The damping rate for $\phi$ is, from equation~\ref{eq:prodrate1}, given by 
\begin{equation}
	\omega \Gamma_\phi = 
	-\Pi^{\phi\phi}_i - \frac{\Pi^{AA}_i ((\Pi^{A\phi}_r)^2 - (\Pi^{A\phi}_i)^2)
	- 2 (\Pi^{AA}_r - m^2) \Pi^{A\phi}_r \Pi^{A\phi}_i}{(\Pi^{AA}_i)^2 + (\Pi^{AA}_r - m^2)^2} ~,
\label{eq:scalardamping}
\end{equation}
Considering a $\phi \bar{e} e$ coupling as an example, this gives
\begin{equation}
	\Gamma_\phi \simeq \frac{g^2}{e^2} k^2 \omega \frac{\omega \sigma_L}{(\omega \sigma_L)^2 + (\omega^2 - \omega_p^2)^2} ~.
	\label{eq:phieenr}
\end{equation}
For $\omega_p < T$, this has a continuum contribution which is approximately
the naive kinetic theory rate. If $m < \omega_p$, then there is also
a resonant contribution, with emission at $\omega \simeq \omega_p$ at a rate
\begin{equation}
	\frac{d\dot N_{\rm prod}}{dV}
	\simeq \frac{1}{4\pi}\frac{g^2}{e^2} k_{\omega_p}^2 \omega_p^2 
	\frac{1}{e^{\omega_p/T} - 1} ~.
			\label{eq:scalarres1}
\end{equation}
Note that, unlike the case for a vector, this resonant contribution
is not suppressed at small $m$. 
This resonant contribution may be larger or smaller than the continuum
one, depending on the properties of the plasma. If $m \ll \omega_p < T$, then
using $\omega_p^2 \simeq n_e e^2 / m_e$, resonant emission gives
\begin{equation}
	Q_{\rm res} \simeq 4 \pi \alpha \alpha_\phi \frac{n_e^2 T}{m_e^2} ~,
\end{equation}
while Compton and bremsstrahlung continuum emission give
(from Appendix~\ref{ap:damp})
\begin{equation}
	Q_{\rm Comp} \simeq \frac{8 \alpha \alpha_\phi}{\pi} n_e \frac{T^4}{m_e^2}
	\quad , \quad
	Q_{\rm brem} \simeq 3 \frac{n_e n_i}{m_e} \sqrt{\frac{T}{m_e}} \alpha^2 \alpha_\phi Z_i^2 ~,
\end{equation}
so
\begin{equation}
	\frac{Q_{\rm res}}{Q_{\rm Comp}} \simeq 
	\frac{\pi^2}{2} \frac{n_e}{T^3}
	\quad, \quad
	\frac{Q_{\rm res}}{Q_{\rm brem}} 
	\simeq
	\frac{4 \pi}{3 \alpha Z_i} \sqrt{\frac{T}{m_e}} ~.
	\label{eq:qresratio}
\end{equation}
We see that resonant production occurs at the same order in $\alpha$ as
Compton production, but for a dense plasma where $n_e > T^3$, the resonant
rate is larger. Bremsstrahlung production is suppressed by $\alpha$ compared
to resonant production, but is enhanced by the inverse electron velocity.
Since electron velocities in stellar cores can be almost relativistic, the resonant contribution can dominate, as we will see in Section~\ref{sec:scalarbounds}.

For relativistic / degenerate electrons, the calculations are slightly
more complicated (see Appendix~\ref{ap:scalar} for details). However,
the overall picture of resonant production being unsuppressed at small
$m$, and being potentially larger than the continuum contribution,
remains. 


\subsubsection{Coupling to nucleons}

In the low-energy SM, we can consider a scalar coupling to
nucleons, $\phi \bar{n} n$ and $\phi \bar{p} p$.
For simplicity, we will take the $n$ and $p$ couplings
to be equal here (this is approximately true for e.g.
a Higgs portal scalar) --- it is simple to extend to
unequal couplings.

The real part of the mixing self-energy is, for non-relativistic nucleons,
\begin{equation}
	\Pi^{\phi L}_r \simeq \sum_i e g Z_i A_i \frac{n_i}{m_i} \frac{k \sqrt{K^2}}{\omega^2}
	\equiv \frac{k \sqrt{K^2}}{\omega^2} \Omega_{eB}^2 ~,
\end{equation}
where $\Omega_{eB}$ is suppressed by $\OO(m_e^2/m_n^2)$ relative to the plasma
frequency.
For a single species of ion, $\Omega_{eB}^2 \simeq (g/e) (m_e/m_n) \omega_p^2$.
The damping rate is again given by equation~\ref{eq:scalardamping}, which 
simplifies to
\begin{equation}
	\omega \Gamma_\phi \simeq (-\Pi_i^{\phi\phi}) + \frac{k^2}{\omega^2} \Omega_{eB}^2 \left(
	\frac{
		2 \omega \Sigma (\omega^2 - \omega_L^2) + \omega \sigma_L \Omega_{eB}^2
		}
	{\omega^2 \sigma_L^2 	+ (\omega^2 - \omega_L^2)^2}\right) ~,
\end{equation}
where we have written $-\Pi^{\phi L}_i = k \sqrt{K^2} \Sigma/\omega$, and taking $\Omega_{eB}^4 \gg \omega^2 \Sigma^2$.
At $\omega \gg \omega_L$, the second term is suppressed by $\Omega_{eB}^2/\omega^2$,
so production is dominated by $-\Pi^{\phi\phi}_i$, as expected.
We also have a resonant contribution to the production rate, given by
\begin{equation}
	\frac{d\dot N}{dV} \simeq \frac{1}{4\pi} \left(\frac{\omega_L}{m} \Pi^{\phi L}_r\right)^2
	\frac{1}{e^{\omega_L/T} - 1} 
			\left| 1- \left. \frac{d\omega_L^2}{d\omega^2}\right|_{\omega_L}\right|^{-1} 
\end{equation}
\begin{equation}
	\simeq \frac{1}{4 \pi} \frac{k_{\omega_p}^2}{\omega_p^2} \Omega_{eB}^4 \frac{1}{e^{\omega_p/T} - 1} ~.
\end{equation}
Both the resonant and non-resonant contributions will, in the case
of equal couplings to protons and neutrons, be suppressed
by $m_e^2 / m_n^2$ relative to the $\phi \bar{e} e$ rates.
As per the $B-L$ case, different ions
have couplings almost proportional to their mass, resulting in a suppression
of ion-ion bremsstrahlung, which would otherwise
contribute at $(m_e/m_n)^{3/2}$. The emissivity relative
to Compton and bremsstrahlung is parametrically the same as in the previous section.


\section{Bounds on weakly-coupled bosons}
\label{sec:bounds}

We now evaluate some of the parametrically-new
production rates discussed in the previous section for the physical
plasmas inside stellar cores, and use these to estimate bounds on the
masses and couplings of new bosons.
We emphasise that our calculations should not be viewed as
precise constraints, which would generally require a detailed numerical
study of the observational data and stellar models.
In addition, we do not aim to be comprehensive --- we choose
a number of new physics scenarios to illustrate our points,
leaving many possibilities for future work.

Following~\cite{Raffelt:1996wa}, there are a number of scenarios
in which we are particularly sensitive to novel energy losses from stars:
\begin{itemize}
	\item \emph{The Sun}. The Sun is not an exceptional star, but it is 
		better-measured than others --- in particular, it is the only
		star whose neutrino emission . With modern solar models,
		novel forms of energy loss can be constrained down
		to around 10\% of the measured solar luminosity~\cite{Gondolo:2008dd,Redondo:2013lna}
		(global modelling using specific energy loss
		profiles can improve this limit to the few percent level~\cite{Vinyoles:2015aba}).
		This gives a limit on the average energy loss per unit mass of
		$\epsilon_{\rm new} \lesssim 0.2 \egs$.
		In the solar core, which is roughly half hydrogen and helium by mass,
		\begin{equation}
		T_{\rm core} \sim 1 \keV \quad , \quad \rho_{\rm core} \sim 150 \gcmt
			\quad , \quad \omega_p \sim 0.3 \keV ~.
		\end{equation}
	\item \emph{Red Giant (RG) cores just before helium ignition}.
		Red Giant cores before the onset of helium fusion are basically
		small, hot white dwarfs at the centre of the enormously-larger
		stellar atmosphere, whose electrons are mostly degenerate.
		Since the core is inert (not undergoing fusion), the
		main energy transfer process is energy loss by neutrino
		emission, $\epsilon_\nu \sim 4 \egs$. Novel
		energy loss processes can be constrained down to around this
		magnitude, 
		$\epsilon_{\rm new} \lesssim 10 \egs$, since a more efficient loss
		process would delay the onset of helium ignition,
		in disagreement with observations that match stellar models
		(e.g.\ \cite{Viaux:2013lha}).
		The conditions in the core just before helium ignition
		are approximately
		\begin{equation}
		T_{\rm core} \sim 10 \keV \quad , \quad \rho_{\rm core} \sim 10^6 \gcmt
			\quad , \quad \mu_e \sim 1.3 m_e \quad , \quad \omega_p \sim 20 \keV ~.
		\end{equation}
	\item \emph{Horizontal branch (HB) star cores during helium-burning}.
		During their helium-burning phase, the energy released by fusion puffs up
		the core of the star, lowering its density and making it non-degenerate.
		The power from helium fusion is $\epsilon_{3 \alpha} \sim 80 \egs$.
		If there are additional energy-loss processes, these cause
		the core to contract, heating it up, enhancing the rate of the helium
		fusion, and so shortening the helium-burning lifetime of the star.
		This lifetime is measured to around the $\sim 10\%$ level, and agrees
		with standard stellar models, so novel energy losses are constrained
		to be $\epsilon_{\rm new} \lesssim 10 \egs$ constraints (e.g.\ \cite{Ayala:2014pea}).
		The core properties are approximately,
		\begin{equation}
		T_{\rm core} \sim 10 \keV \quad , \quad \rho_{\rm core} \sim 10^4 \gcmt
			\quad , \quad \omega_p \sim 2 \keV ~.
		\end{equation}
	\item \emph{Core-collapse supernovae (SN)}. In the first few seconds of
		a core-collapse supernova, the core is basically a very hot
		proto-neutron star, with degenerate electrons and almost-degenerate
		nucleons. It is sufficiently hot and dense to be optically thick
		even to neutrinos. The dominant energy loss
		mechanisms is via the outwards diffusion of neutrinos, which
		occurs over a $\sim 10 \second$ timescale --- the resulting
		neutrino burst was measured for SN1987A~\cite{Raffelt:1996wa}, with properties that
		agreed with SN model predictions. In order for a new energy-loss
		mechanism not to have disrupted the SN1987A neutrino burst, 
		the averaged energy loss rate would have to have
		been less than that from the neutrinos,
		$\epsilon_{\rm new} \lesssim 10^{19} \egs$~\cite{Raffelt:1996wa}.
		\begin{equation}
			T_{\rm core} \sim 30-60 \MeV \quad , \quad \rho_{\rm core} \sim 3 \times 10^{14} \gcmt
			\quad , \quad \mu_e \sim 350 \MeV
			\quad , \quad \omega_p \sim 20 \MeV ~.
		\end{equation}
		As we will see, for particle emission processes that do not depend on high powers
		of the temperature or density, stellar limits are generally more
		constraining than SN limits for masses $\lesssim 50 \keV$. However, SN
  		limits apply up to masses $\sim 100 \MeV$.

		It should be emphasised that there is still significant
		uncertainty regarding the behaviour of core-collapse
		supernovae~\cite{Muller:2016izw}, with conflicting models in the literature
		of even the very basic energy source involved~\cite{Blum:2016afe}.
		As a result, SN constraints on new light particles should
		be viewed as order-of-magnitude estimates at best.

	\item \emph{Very degenerate stellar remnants: white dwarfs (WDs) and neutron
		stars (NSs)}. These have the desirable properties of fairly
		inefficient SM energy-loss mechanisms (either neutrino emission, or
		surface photon emission), as well as reasonable temperature 
		and high density compared to the aforementioned stellar core situations.
		However, the emission of new particles in scattering process
		is generally highly suppressed by Pauli-blocking.
		As a result, WDs can only impose similar constraints
		to those from stars in some cases,
		while NSs do not seem to be competitive~\cite{Raffelt:1996wa}.
		For resonant emission, the situation is ever worse: since
		these highly degenerate systems have plasma frequencies
		$m_T, \omega_L \gg T$, emission at resonant frequencies
		is heavily Boltzmann-suppressed, so is never significant.
\end{itemize}


\subsection{Emission and trapping}
\label{sec:emtrap}

The production calculations in Section~\ref{sec:thermalprod}
looked at the production
rate per unit volume for a weakly-coupled in-medium mode,
assuming a uniform, time-independent thermal medium.
For stellar cooling bounds, we are interested in the 
energy flux that escapes the core --- that is, in the energy
transport between regions with different medium properties.
However, as long as medium properties vary slowly
enough, then a weakly-coupled mode emitted from one
region will propagate adiabatically into 
the weakly-coupled mode for the new region
(if the weakly-coupled and strongly-coupled modes
go through level-crossing during the propagation, there
will be resonant conversion, which needs to be taken
into account).

Consequently, as long as absorption (and resonant conversion)
during the course of the propagation is small, almost
all of the energy emitted in weakly-coupled excitations
at the centre is transported to weakly-coupled excitations
further out. When the whole star is optically thin to weakly-coupled
excitations, energy is lost to infinity, and the total energy
loss rate is obtained by integrating over the per-volume production rate.

If weakly-coupled excitations are not able to free-stream through the
whole star, then they contribute to energy transport within the star,
and let the outer layers of the star lose energy to infinity (once the
optical depth becomes low). Since our models
of stellar structure fit observations fairly well, there is generally no room
for a whole additional degree of freedom to contribute to radiative heat
transport, unless that degree of freedom is heavy enough that its
Boltzmann-suppressed abundance is too small to contribute
significantly~\cite{Raffelt:1996wa}. We will see
in Sections~\ref{sec:blbounds} and~\ref{sec:scalarbounds} how this
opens up some extra parameter space at high masses and couplings,
for stellar bounds on new particles.
As discussed in Section~\ref{sec:sntrap}, numerical simulations of supernovae
also seem to indicate that energy transport by particles
with mean free path significantly larger than that of neutrinos
is not compatible with SN1987A observations, allowing similar
energy-transport bounds to be placed.


\subsection{Dark photon production in supernovae}
\label{sec:dpsn}

The uncertainty surrounding the behaviour of core-collapse supernovae is
an obstacle to using detailed explosion models to constrain new physics.
As mentioned in the Introduction, our focus in this paper is on
highlighting the physical consequences of plasma effects, rather than
attempting comprehensive and robust stellar modelling. Accordingly,
we adopt a single representative supernova
model (from~\cite{Keil:1994sm}, which is similar
to others found in the literature, see e.g.\ the review~\cite{Janka:2012wk}),
and impose the simple `Raffelt bound' approximation on
the energy loss from new particles.
That is, we demand that the average energy loss per unit mass
from the core via new particles is $\lesssim 10^{19} \egs$~\cite{Raffelt:1996wa}
--- in a variety of simulations, this is within a factor $\sim 10$
of the critical energy loss rate at which the neutrino signal
would be detectably modified~\cite{Hanhart:2001fx}.
Thus, the constraints we derive should not be considered as
robust parameter space exclusions. However, as we will see,
plasma mixing can lead to orders-of-magnitude differences in the
production and absorption rates, compared to a kinetic theory
calculation~\cite{Dent:2012mx,Kazanas:2014mca,Rrapaj:2015wgs}.
Such effects will therefore be very important to include in any
eventual robust study, and our calculations indicate how such exclusion
bounds will vary with dark photon mass.

In the supernova environment, the real part of the photon self-energy
is dominantly generated by electrons\footnote{
	The proton contribution is suppressed, since protons are non-relativistic
	in the SN; the neutron-neutron
contribution vanishes in the limit that neutrons can be treated as free,
and remains suppressed after allowing for interactions in a typical SN
environment~\cite{Kopf:1997mv}}, the form of which is discussed
in Section~\ref{sec:smplasma} and Appendix~\ref{ap:photonpi}. 
The imaginary parts of the photon self
energy gets contributions from various processes, and the large rate of
nucleon-nucleon interactions means that neutron--proton bremsstrahlung
is dominant. Following~\cite{Rrapaj:2015wgs}, we can approximate 
the bremsstrahlung production rate by relating it to proton-neutron
scattering data, using the soft approximation for photon emission.
This is only strictly correct in the limit that the emitted energy is small compared
to the kinetic energy of the collision, but from~\cite{Rrapaj:2015wgs},
we expect that the overall error from this approximation 
will be comparable to the other uncertainties in our calculation.
Further details are given in Appendix~\ref{ap:reddy}.
For some supernova models, the SM photon oscillations may become
broad resonances in the supernova core, in which case the split of dark photon production
into continuum and resonant contributions loses its applicability.

\begin{figure}
	\begin{center}
		\includegraphics[width=.8\textwidth]{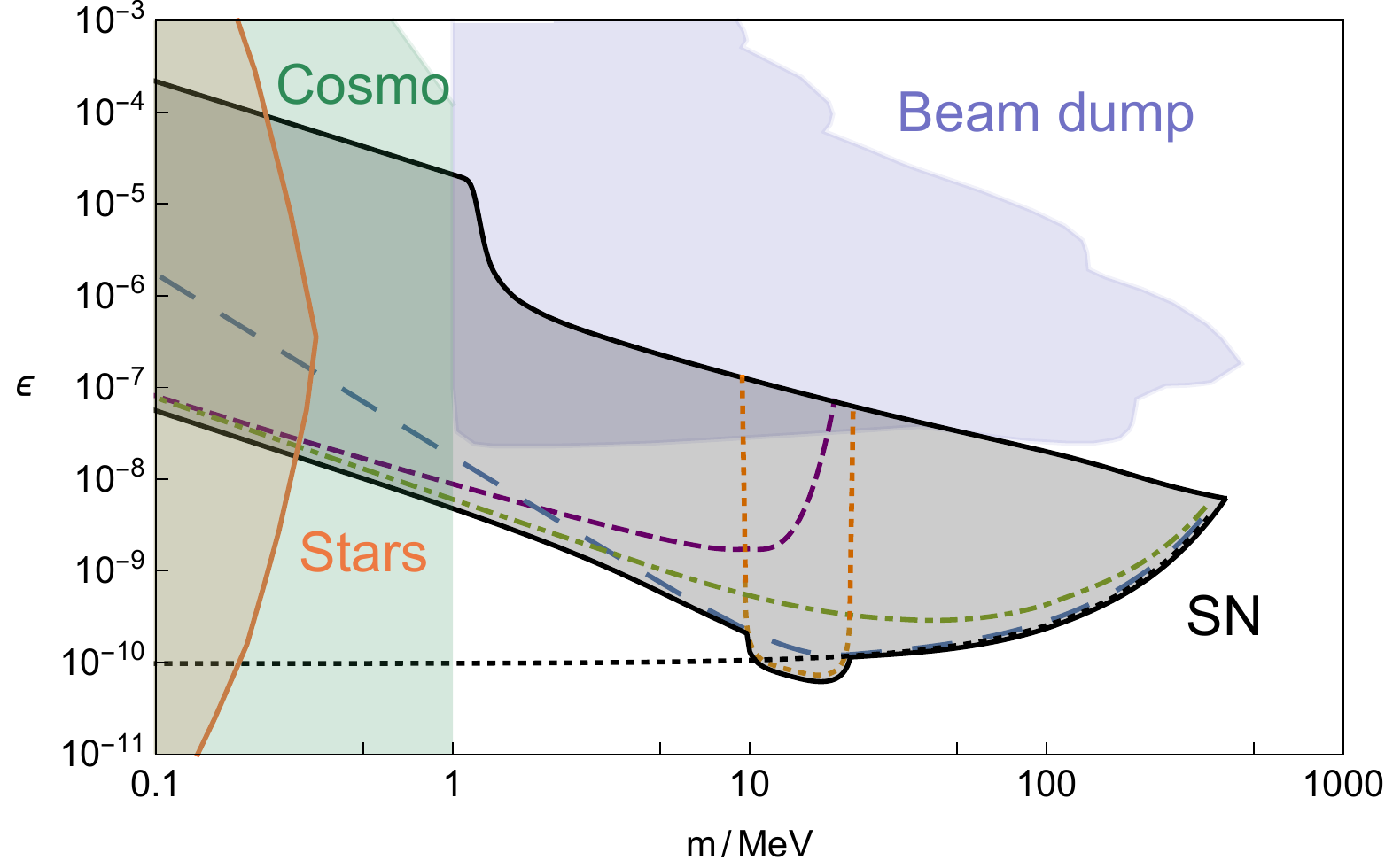}
	\end{center}
	\caption{Constraints on dark photon couplings
	from SN1987A observations, for a particular
	supernova model as described in the text. 
	The solid black region indicates the overall constraints.
	Subcomponents of the total emission are indicated;
	dotted (orange), resonant transverse emission,
	long-dashed (blue), continuum transverse emission,
	dot-dashed (green), continuum longitudinal emission,
	dashed (purple), resonant longitudinal emission.
	The limit on $\epsilon$ that would
be obtained from a kinetic theory calculation is shown in dotted black. 
	Assuming
purely SM decays of the dark photon, the parameter space above the shaded region is not
	excluded by the energy loss constraint we impose because of absorption and/or decays to electron-positron pairs (at masses $\gtrsim 2 \MeV$) (although energy transport
	via the dark photon may mean that the neutrino signal is still observably
	shortened --- see Section~\ref{sec:sntrap}). Under the same assumption,
 beam dump experiments exclude the region shaded in
blue~\cite{Essig:2013lka}. The orange shaded region is excluded
	by stellar cooling bounds~\cite{An:2013yfc,Redondo:2013lna} (where
	we have also incorporated the effects of trapping inside
	the star, as per Section~\ref{sec:emtrap}), while in models without hidden sector decays masses
	below $\sim \MeV$ will be constrained by cosmological observations
	(green, \cite{Redondo:2008ec} --- the small gap at large mixing
	will very probably be closed by beam dump experiments such as~\cite{Banerjee:2016tad}).
	It should be stressed that models of core-collapse supernovae are
	very uncertain, and limits such as those derived here,
	which assume a particular SN model~\cite{Keil:1994sm},
	should be treated as order-of-magnitude estimates at best.
	}
	\label{fig:sn1}
\end{figure}

In Figure \ref{fig:sn1} we plot the bounds obtained for our SN model after 
including the mixing effects discussed in
Section~\ref{sec:dp} --- the bounds that would be obtained
neglecting such
effects are also shown.\footnote{
	Our results are compatible with~\cite{Chang:2016ntp}, which
	considers a range 
	of SN profiles, finding roughly an order of magnitude variation in the bounds obtained. Our SN profile is close to those denoted ``Fischer'' in~\cite{Chang:2016ntp}, and leads to similar constraints.}
At masses below the typical plasma frequency, transverse
continuum emission has the expected  $\epsilon \sim 1/m^2$ scaling. Meanwhile, 
longitudinal continuum and resonant emission both scale as $\epsilon \sim 1/m$.  
At masses around the plasma frequency, resonant transverse emission
is possible, and leads to a stronger constraint, while at higher masses continuum
emission of longitudinal and transverse modes is close to the kinetic theory prediction. 

For dark photon masses below approximately $0.1 \MeV$, bounds from
stellar cooling become dominant, also shown in Figure~\ref{fig:sn1}.
Production in these environments is
less suppressed since the plasma frequency is lower, and they are better
modelled and measured than supernovae.
For masses below $2m_e$ there
are also strong constraints from cosmology --- a small dark
photon abundance is produced in the early universe,
which then decays to three photons, which would be visible
in the observed gamma ray background~\cite{Redondo:2008ec}
(or, for sufficiently early decays, through effects on the CMB
or BBN). 
Such constraints could be relaxed by allowing the dark photon to decay
to light hidden sector states, though the viability of this possibility
is model dependent.
There are also cosmological constraints for $m > 2 m_e$
mixings~\cite{Fradette:2014sza,Berger:2016vxi}, but these constrain significantly
smaller mixings than the supernova cooling limits.

In addition to affecting the neutrino signal through energy loss
from the SN core, new particles that decay to SM could have
other experimentally-visible signatures in SN.
\cite{Kazanas:2014mca} considers some of these signatures for a dark
photon, including gamma radiation from the injection of $e^+ e^-$
pairs when dark photons decay slightly outside the star.
However, they use the kinetic theory calculation for dark photon production,
so their constraints would need re-calculating with the correct
production and absorption rates. We leave this to future work.

\subsubsection{Trapping and decay}
\label{sec:sntrap}

If $\epsilon$ is large enough, then for the SN core there is some
(frequency-dependent) radius within which dark photon radiation
gets reprocessed into nearly black-body radiation.\footnote{
	Along a given `line of sight' out of the core,
	the occupation number $f$ of a weakly-coupled state
	propagating in that direction
	will evolve as
	$v \, df/dx = - f \Gamma_{\rm abs} + (1 + f) \Gamma_{\rm prod}
	= \Gamma_{\rm prod} - f \Gamma_{\rm damp}$
	where $v$ is the velocity of the weakly-coupled state.
	So, for intervals over which the `damping depth'  $\tau_d \equiv \int dx \, \Gamma_{\rm damp}/v$
	is large, the occupation number converges to the black-body value 
	$f_B = \Gamma_{\rm prod}/\Gamma_{\rm damp}$, while
	for intervals over which $\tau_d \ll 1$, we simply
	integrate the production rate along the line.
	For frequencies $\omega \gtrsim T$, we have $\Gamma_{\rm damp}
	\simeq \Gamma_{\rm abs}$, so the damping depth
	is approximately the optical depth. For $\omega \ll T$,
	$\Gamma_{\rm damp} \simeq (\omega/T) \Gamma_{\rm abs}$,
	so the damping depth is significantly smaller than the optical depth.
	\label{foot:depth}
	}
The core will then lose energy, at a given frequency, approximately
like a black-body sphere of that radius (with surface temperature
the medium temperature at that radius), along with volume emission from the region
outside. If these black-body radii are large enough, and correspond to small
enough medium temperatures, then the overall energy loss rate via
the dark photon will become small compared to the neutrino rate.

We make the assumption that, to significantly affect the neutrino
signals, there needs to be energy loss greater than
the Raffelt bound from the part of the core at temperature $\gtrsim 1.5
\MeV$~\cite{Rrapaj:2015wgs}. As per Section~\ref{sec:emtrap},
it is likely that provided the dark photon mean free path is greater than
of neutrinos in the core, the increased efficiency of energy transport
would observably shorten the neutrino signal~\cite{Dolgov:2000pj,Keil:1994sm,Burrows:1990pk}. Imposing this condition would allow dark photon coupling roughly
an order of magnitude stronger than those constrained by the energy loss
bound to be ruled out; however, detailed simulations would be required to
obtain quantiative bounds. Since such strong couplings are already ruled
out by beam dump experiments, we only show the energy loss bounds
in Figure~\ref{fig:sn1}.

 For the
supernova profile we use, the $T= 1.5
\MeV$ boundary is at a radius of approximately
$14.5 \, {\rm km}$, by which point the electron chemical potential has
dropped significantly, to $\sim \MeV$. 
This corresponds to the typical region from which neutrinos can free
stream out of the supernova. 
Note however, it might be that energy has
to escape to much larger distances, before the dark photons suppress neutrino productions significantly~\cite{Chang:2016ntp}.

Up to such radii, continuum absorption of dark photons
is dominated by inverse proton-neutron bremsstrahlung,
expressions for which are given in Appendix~\ref{ap:reddy}
(at much larger radii, inverse Compton scattering off electrons
may become important). The $\omega^{-3}$ dependence of the bremsstrahlung
absorption rate means that low-frequency dark photons are trapped more
efficiently than higher-frequency ones, and so the frequency
dependence of the effective black-body radius is important.
For dark photon masses $\gtrsim 100 \keV$, continuum absorption of
transverse dark photons is more efficient than longitudinal continuum
absorption, because of the small plasma frequency in the outer part
of the supernova core. For small enough
dark photon masses, transverse absorption will
be suppressed by the expected $m^4/\omega_p^4$ factor,
so transverse photons will escape most easily,
but such small masses are constrained by stellar
cooling observations. Longitudinal dark photons can be resonantly
absorbed if there is a shell where the supernova conditions are such
that $\left|\omega^2 - \omega_L^2\right|< \omega \sigma_L$ is satisfied,
and if present this shell is typically optically thick for
$\epsilon$ at the trapping bound. Meanwhile,
resonant reabsorption of transverse modes occurs for dark photon masses
around the typical plasma frequencies.

If the dark photons decay to electrons
and positrons
before escaping the neutrino emission region, this will also prevent
efficient energy loss (although as for absorption, the neutrino signal may still be observably shortened). In the centre of the supernova, decay to electrons
is blocked by the high chemical potential, unless the mass of the dark
photon is above twice the effective electron mass, $m_{\rm eff} \simeq
15 \MeV$. However, towards the outside, the chemical potential drops
sharply, and the effective electron mass is close to its vacuum value.
In Appendix~\ref{ap:ep} we give explicit formula for the decay rate.

The effect on the allowed values of $\epsilon$ is plotted in Figure
\ref{fig:sn1}. For dark photon masses $m\gtrsim
2m_e$, decays dominate the absorption, although this is sensitive to the details of the supernova profile.
If the dark photon can decay to lighter hidden sector states, which
interact sufficiently weakly with the supernova medium that they can
escape, decays and absorption would be reduced and larger values of
$\epsilon$ constrained. This could also relax limits from beam dump
experiments, and, as mentioned, alter cosmological constraints. The
interplay of these effects would be interesting to study in the future.


\subsection{$B-L$ vector bounds}
\label{sec:blbounds}

\begin{figure}
	\begin{center}
		\includegraphics[width=.8\textwidth]{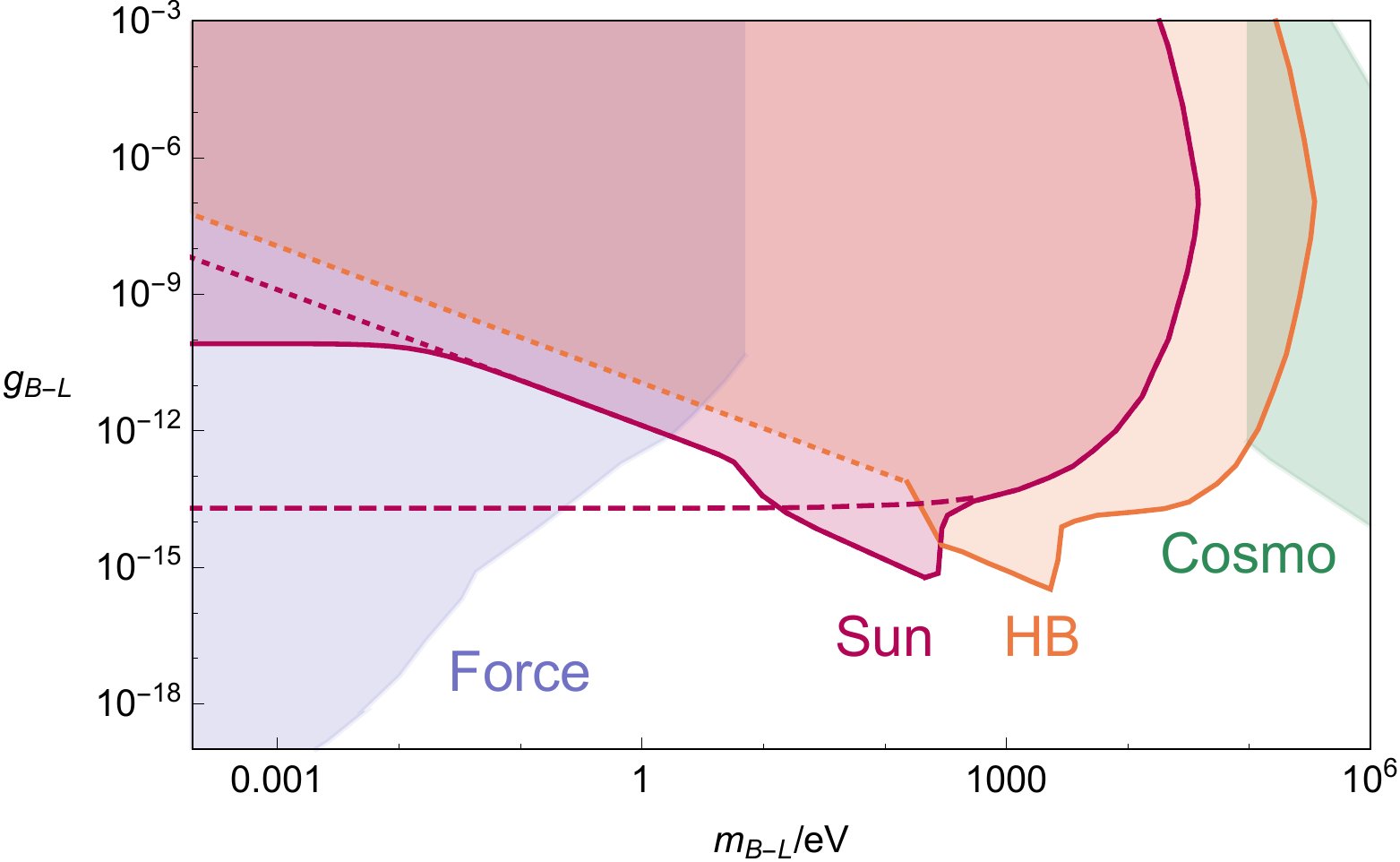}
	\end{center}
	\caption{
		Constraints on the coupling of a $B-L$ vector, from force tests (blue, \cite{Decca:2007jq,Sushkov:2011zz,Geraci:2008hb,Kapner:2006si,Hoskins:1985tn}),
		stellar cooling (see text) and cosmological observations (green, \cite{Redondo:2008ec}).
		The solar bounds (red) take into account the mass-independent
		part of the production rate discussed in Section~\ref{sec:blprod},
		shown by the solid line --- the dotted line shows
		the constraints that would arise if the coupling to neutrons
		were neglected.
		As explained in the text, the value of the mass-independent
		contribution to the production shown here is an order-of-magnitude estimate.
		The HB star bounds (orange) do not take into account the coupling to neutrons,
		so are shown dotted at low masses, where the neutron
		contribution will be important.
		The dashed red line indicates the bounds that would be obtained
		from a naive `kinetic theory' calculation, which ignored in-medium mixing effects.
	}
	\label{fig:bl1}
\end{figure}

From Section~\ref{sec:blprod}, B-L production in a dilute,
non-relativistic plasma should be well-approximated by the dark photon
production rate, plus a mass-independent production rate that is suppressed by at
least $(m_e/m_n)^{3/2}$. As discussed in footnote~\ref{foot:blprod},
the proton velocity in the Sun is small enough that
the extra $\sim v/\alpha$ suppression of ion-ion bremsstrahlung
makes it numerically comparable to the $(m_e/m_n)^2$ contribution
from electron-ion bremsstrahlung. The plasma screening length inside
the solar core is also small enough that it may affect
the ion-ion bremsstrahlung calculation. 
We have not treated all of these contributions precisely,
so our numerical value for the mass-independent production rate
should be treated as an order-of-magnitude estimate.
For the Sun, where the core
is roughly half hydrogen and half $^4$He by mass, we obtain
the bounds shown in in Figure~\ref{fig:bl1}.
There will also be low-mass bounds from HB
and RG stars --- however, since these are unlikely
to rule out significant areas of parameter space not already constrained
by force bounds, we restrict ourselves to the solar calculation as an illustration.
Figure~\ref{fig:bl1}
also shows the trapping bounds for emission from the Sun and from HB stars,
as discussed in Section~\ref{sec:emtrap} --- these constraints translate directly
to the dark photon case, in the high-mass region.

Comparing to the existing literature, 
stellar bounds on massive $B-L$ vectors have been considered
in~\cite{Carlson:1986cu,Harnik:2012ni,Heeck:2014zfa}. \cite{Carlson:1986cu}
performs the naive kinetic theory calculation, ignoring plasma mixing
effects.
\cite{Harnik:2012ni, Heeck:2014zfa} treat the production of on-shell vectors
as being the same as for dark photons, which ignores the 
mass-independent production rate from non-hydrogen nuclei
calculated in Section~\ref{sec:blprod}.
\cite{Harnik:2012ni, Heeck:2014zfa} also claim that there is a strong
bound from the enhanced emission of SM neutrinos.
However, the $B-L$ mediated emission of neutrinos has
rate $\sim g_{B-L}^4$, which is extremely small, while the rates
in the aforementioned papers appear to be $\sim g_{B-L}^2$.

Figure~\ref{fig:bl1} also shows the bounds on a light $B-L$ vector
from fifth-force tests~\cite{Decca:2007jq,Sushkov:2011zz,Geraci:2008hb,Kapner:2006si,Hoskins:1985tn}, which are more constraining than stellar
cooling bounds for masses $\lesssim \eV$. It also indicates the cosmological
bounds arising from late-time three-photon decays of a $B-L$ population,
produced in the early-universe plasma. These can be derived directly from
the dark photon bounds~\cite{Redondo:2008ec}, since the dominant
contribution to early-universe production happens
around electron-positron freeze-out, when only the electron coupling is important.
The cosmological bounds depend on the $B-L$ vector
not having hidden sector decays.
Another possible probe of $B-L$ vectors or dark photons
would be the direct detection of weakly-coupled states
emitted from the Sun, in dark matter experiments
on Earth~\cite{An:2013yua,An:2014twa}. 

As is the case for dark photons, there will also be supernova cooling
bounds on $B-L$ vectors of mass $\lesssim$ a few hundred $\MeV$.
Since neutrons play an important role in a supernova core, we expect
the constraints to be rather different from those in the dark photon
case. We leave an analysis of these to future work.

The importance of plasma mixing effects does not just apply to
$B-L$ vectors or dark photons, but to any new vector coupling to the
SM. For example, the bounds on a new vector coupling to baryon number
derived in~\cite{Grifols:1988fv,Rrapaj:2015wgs} did not consider mixing effects, so
would need to recomputed with those taken into account.
Again, we would expect kinetic-theory-like emission
at energies above the plasma frequencies, some enhancement from
resonant emission at masses around the plasma frequencies,
and then some suppression due to interference effects
at lower masses.


\subsection{$\phi \bar{f} f$ scalar bounds}
\label{sec:scalarbounds}

As discussed in Section~\ref{sec:scalarprod}, the production
rate for a scalar that couples coherently to SM plasmons
has a resonant contribution for $m < \omega_p$, and a continuum
contribution that is roughly similar to the naive kinetic
theory value. The resonant contribution is lower-order
in $\alpha$ than the continuum bremsstrahlung rate, and
is expected to be larger for plasmas with large electron
velocities.

\subsubsection{Coupling to electrons}

Using the non-relativistic resonant production rate from
equation~\ref{eq:scalarres1}, and taking a sample HB star model
from~\cite{Dearborn:1989he},
imposing the rough constraint $\epsilon \lesssim 10 \egs$ gives
\begin{equation}
	\alpha_{\phi \bar{e} e} \lesssim 8 \times 10^{-31} ~,
\end{equation}
for $m \ll \keV$, where $\alpha_{\phi \bar{e} e} \equiv 
g_{\phi\bar{e} e}^2 / (4\pi)$. For comparison, the kinetic theory production
rate would give $\alpha_{\phi \bar{e} e} \lesssim 1.5 \times 10^{-29}$
(in agreement with the estimate from~\cite{Raffelt:1996wa}).
Since the continuum production is dominated by $^4$He bremsstrahlung, 
we expect the ratio of resonant to transverse power per volume to be
given by equation~\ref{eq:qresratio},
\begin{equation}
	\frac{Q_{\rm res}}{Q_{\rm brem}} 
	\simeq
	\frac{4 \pi}{3 \alpha Z_i} \sqrt{\frac{T}{m_e}}
	= 37 \sqrt{\frac{T}{10^8 \kelvin}} ~,
\end{equation}
Since our HB model has $\langle \sqrt{T/10^8 \kelvin} \rangle \simeq 0.5$
averaged over the core,
this gives the correct ratio of $\sim 20$ between the continuum and resonant
constraints.

\begin{figure}
	\begin{center}
		\includegraphics[width=.49\textwidth]{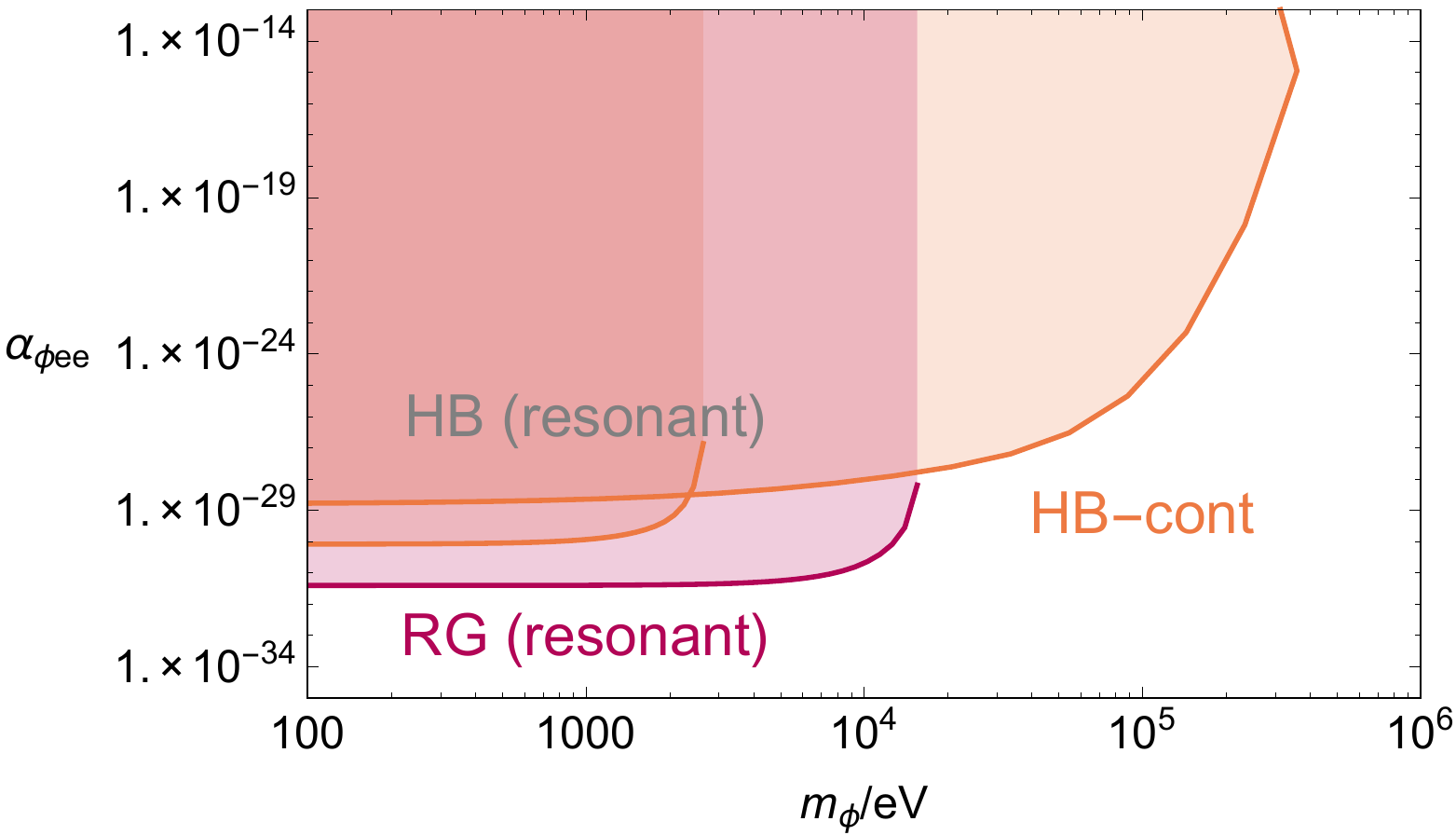}
		\includegraphics[width=.49\textwidth]{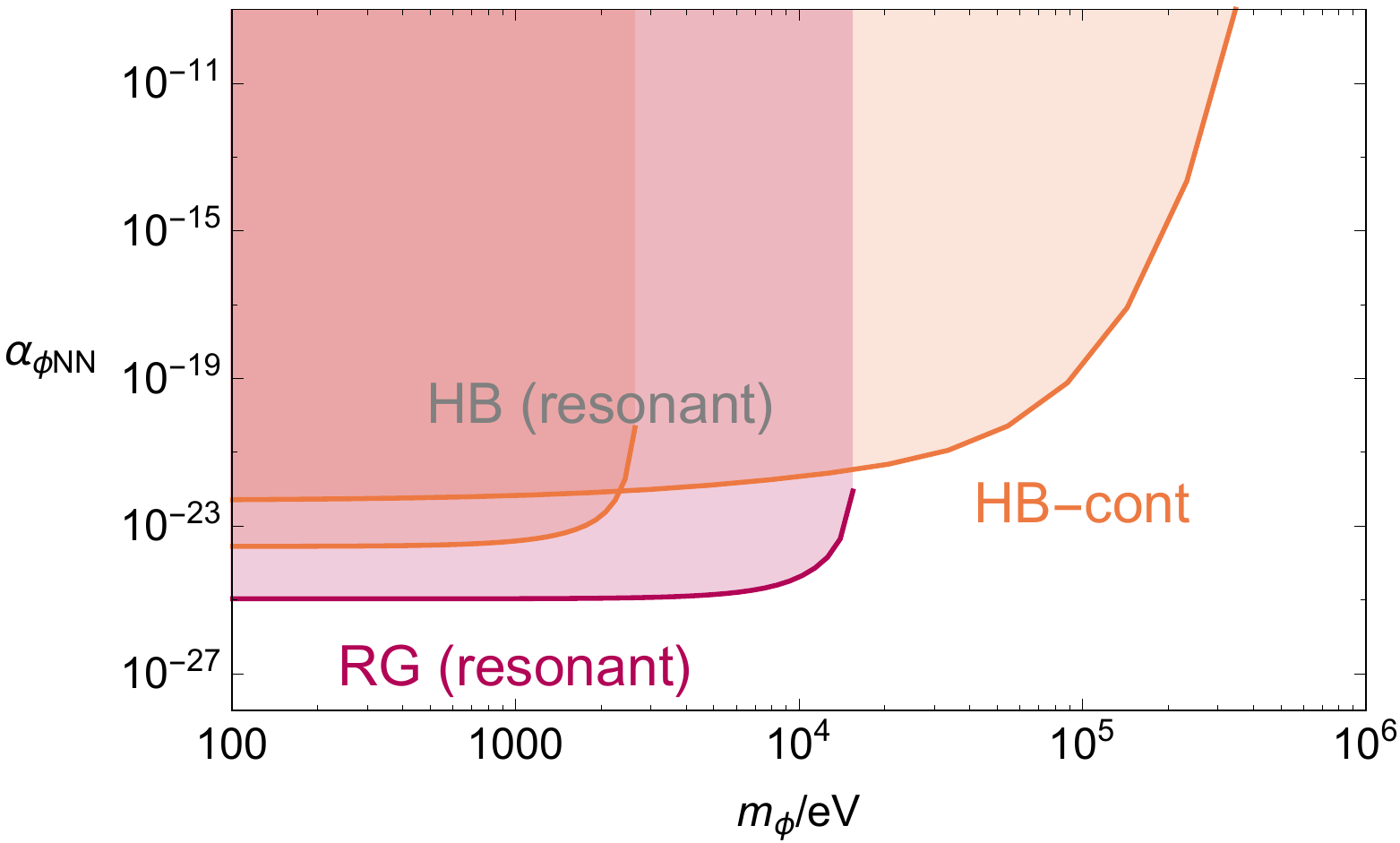}
	\end{center}
	\caption{
		Stellar cooling constraints on a scalar $\phi$ with a
		$\phi \bar{e} e$ coupling (left), or a $\phi \bar{N} N$ coupling to nucleons
		(right), where we assume that $\phi$ couples equally to protons and nucleons.
		Both the continuum and resonant limits continue to very small $\phi$ masses.
		Continuum production of $\phi$ in HB stars gives approximately the same
		energy loss as a naive kinetic theory calculation.
		There will also be continuum production from RG cores,
		but this is more complicated to calculate due to electron degeneracy,
		so we leave it to future work (we expect it to be sub-dominant
		to resonant production at small masses).
		At high masses and couplings, the contribution of $\phi$
		to energy transport in the star becomes small enough
		that it is not constrained by observations (Section~\ref{sec:emtrap}),
		as shown the in left-hand plot (the trapping constraints occur at larger
		$\alpha$ than shown in the right-hand plot).
		In a full model, there will also be other constraints on these
		parameter spaces, including force bounds and cosmological
		observations (for example, Figure~\ref{fig:hp1} shows the constraints
		on a Higgs portal scalar model).
	}
	\label{fig:eenn}
\end{figure}

For a RG core, where the electrons are degenerate, we can use
the formulae from Appendix~\ref{ap:scalar} to calculate the resonant
production rate. Using a sample RG core model from the MESA
package~\cite{Paxton:2010ji}, and imposing $\epsilon
\lesssim 10 \egs$, we obtain
\begin{equation}
	\alpha_{\phi \bar{e} e} \lesssim 4 \times 10^{-32} ~,
\end{equation}
for $m \ll 10 \keV$. Calculating the continuum production rate
in this case is more difficult, due to the electron degeneracy,
but we expect resonant production to dominate at small masses.
To summarise, for small scalar masses, we have the constraints
\begin{equation}
	\alpha_{\phi \bar{e} e} \lesssim 
	\begin{cases}
		1.5 \times 10^{-29}  &\mbox{(HB continuum production)} \\
		8 \times 10^{-31}  &\mbox{(HB star resonant production)} \\
		4 \times 10^{-32}  &\mbox{(RG core resonant production)} ~. \\
	\end{cases}
\end{equation}
The corresponding mass-dependent constraints are shown in Figure~\ref{fig:eenn}.

A calculation similar to that performed in Section~\ref{sec:dpsn}
could be done to find SN bounds on the $\phi \bar{e} e$ coupling.
We are not aware of any existing calculation in the literature,
and we leave such a treatment to future work.

\subsubsection{Coupling to nuclei}

Performing the analogous calculations for a scalar coupling to nucleons,
assuming that we couple equally to protons and neutrons, we obtain
\begin{equation}
	\alpha_{\phi \bar{N} N} \lesssim 
	\begin{cases}
		5 \times 10^{-23}  &\mbox{(HB continuum production)} \\
		3 \times 10^{-24}  &\mbox{(HB star resonant production)} \\
		1 \times 10^{-25}  &\mbox{(RG core resonant production)} ~,\\
	\end{cases}
\end{equation}
for small scalar masses (note that our continuum production bound is slightly tighter than
that given in~\cite{Raffelt:1996wa}, since only Compton scattering is
considered there, and electron-ion bremsstrahlung
production is dominant for low-mass scalars). We again obtain
a ratio of $\sim 20$ between resonant and continuum production for
HB stars, and a significantly stronger bound from RGs.
Figure~\ref{fig:eenn} shows the mass-dependent bounds.


\subsubsection{Higgs portal scalar}
\label{sec:hpphi}

\begin{figure}
	\begin{center}
		\includegraphics[width=.8\textwidth]{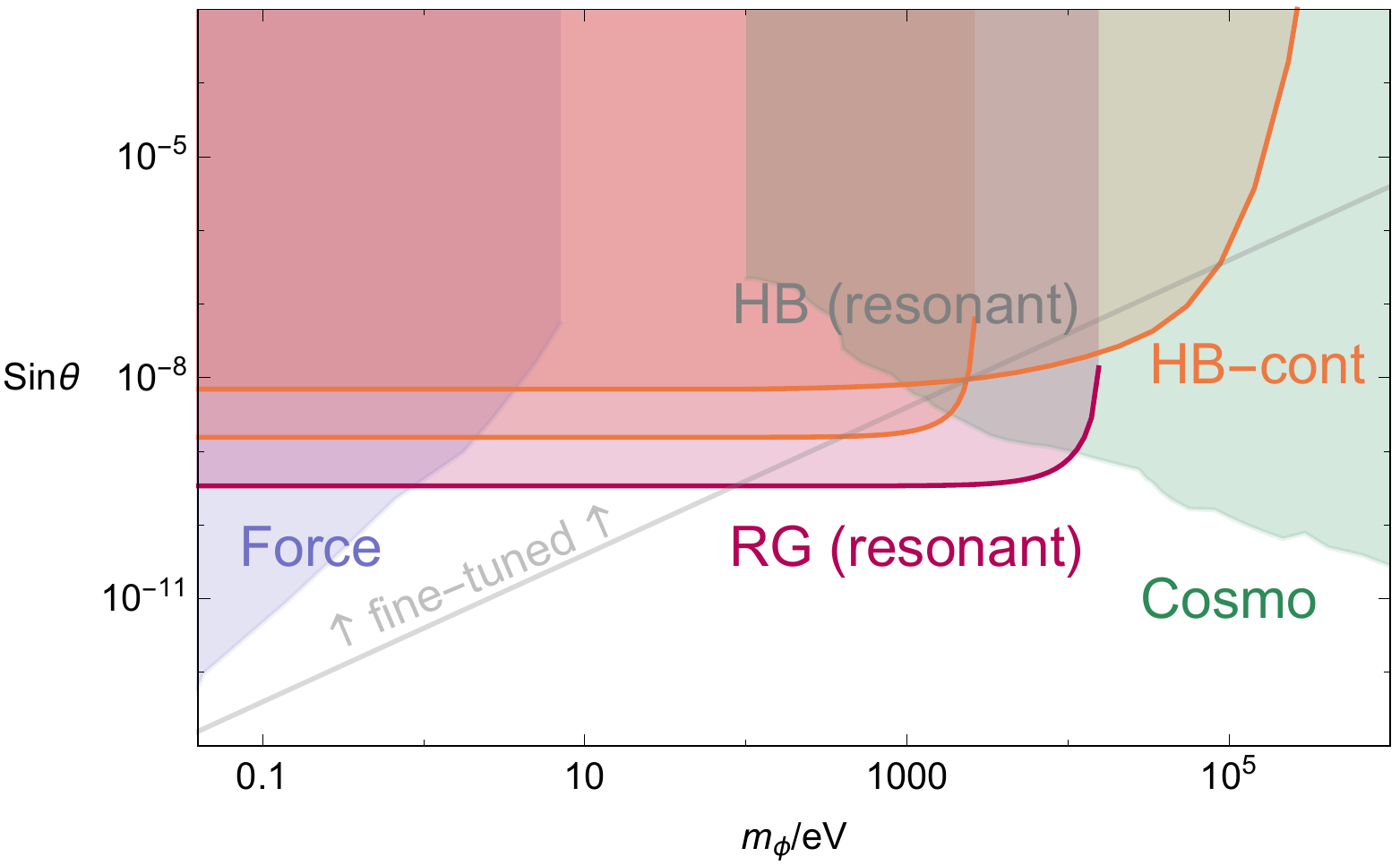}
	\end{center}
	\caption{
		Constraints on a scalar $\phi$ with a Higgs portal coupling (Section~\ref{sec:hpphi}),
		where $\sin\theta$ is the mixing with the SM Higgs.
		Force bounds~\cite{Decca:2007jq,Sushkov:2011zz,Geraci:2008hb,Kapner:2006si,Hoskins:1985tn}
		are in blue, stellar cooling bounds (see text) in red and orange,
		and cosmological constraints in green.
		The latter arise from a freeze-in relic abundance decaying around
		the present day, contributing to the observed
		photon background~\cite{Flacke:2016szy}
		(we show these bounds as extending up to couplings
		at which $\phi$ would decay at BBN time, though it would require
		further work to verify whether the constraints are valid there).
		Since most of the nucleon mass comes from QCD, rather than
		from the Higgs VEV, the ratio of nucleon to electron $\phi$ couplings
		is smaller than the nucleon-electron mass ratio.
		The electron coupling therefore dominates the $\phi$-SM mixing,
		and thus the resonant $\phi$ production, as well as dominating
		the continuum production.
		The stellar cooling bounds are therefore the appropriate rescaling 
		of the $\phi \bar{e} e$ bounds show in the left-hand panel of
		Figure~\ref{fig:eenn}, and likewise continue to very small $\phi$
		masses.
		The grey line shows the approximate fine-tuning bound $\sin\theta
		\lesssim m_\phi / v_{\rm EW}$~\cite{Graham:2015ifn}, demonstrating that our new stellar constraints
		are able to probe natural parameter space.
	}
	\label{fig:hp1}
\end{figure}

The $|H|^2$ operator is the unique renormalisable
portal through which a neutral scalar can be coupled to the SM.
In the low-energy SM, its effects can be parameterised by
a mixing angle $\sin\theta$, with the scalar $\phi$ coupling to fermions
as $(m_f/v)\sin\theta \phi \bar{f} f$, where $v \simeq 246 \GeV$
is the electroweak vev. This gives a coupling to
electrons of $g_{\phi \bar{e} e} = (m_e/v) \sin\theta = 2 \times 10^{-6}
\sin\theta$, and a coupling to nucleons of 
$g_{\phi \bar{N} N} \simeq 8 \times 10^{-4} \sin\theta$~\cite{Dawson:1989gr} 
(this is approximately the same for neutrons and protons, since
most of the coupling does not come from the valence quarks).
Since $g_{\phi \bar{e} e} / g_{\phi \bar{N} N} \simeq 2\times 10^{-3}
> m_e/m_n \simeq 5 \times 10^{-4}$ --- as expected,
since most of the nucleon mass comes from QCD, not from the Higgs VEV --- the $\phi \bar{e} e$ coupling
dominates the $\phi$ production rate in stellar cores,
resulting in the stellar cooling bounds shown
in Figure~\ref{fig:hp1}.
An interesting point is that the enhanced stellar cooling
constraints
are able to probe potentially natural parameter space for
a Higgs portal scalar, as indicated in Figure~\ref{fig:hp1}.

Comparing these to other constraints, the stellar cooling bounds are 
more stringent than fifth-force constraints at masses $\gtrsim 0.2 \eV$.
Having a UV complete model also enables us to consistently compare
the stellar bounds to cosmological constraints (assuming that the universe was
reheated above the electroweak scale, and that there was no cosmologically-important
new physics below the EW scale).
\cite{Flacke:2016szy} estimates the early-universe $\phi$ production,
which is dominated by temperatures around the electroweak scale,
and then constrains the later energy injection from $\phi \rightarrow
\gamma \gamma$ decays. This gives the bounds shown in Figure~\ref{fig:hp1},
indicating that the stellar bounds are probably the most constraining up
to masses $\sim 10 \keV$. We leave an improved calculation of early-universe
$\phi$ production, incorporating plasma effects, to future work.
It would also be interesting to calculate the minimal bounds,
assuming only that the reheating temperature is $\gtrsim \MeV$.

\section{Discussion}
\label{sec:discuss}

In this work, we have demonstrated how, in the medium
of a SM thermal bath, plasma effects can introduce
an effective in-medium mixing between hypothetical new, weakly-coupled bosons
and SM photons (if the new bosons couple coherently to SM plasma oscillations).
Such mixing affects the production of the new bosons
by allowing resonant production from SM photons,
and also by allowing cancellation between new boson and photon
emission amplitudes. This can make a parametric difference to
production rates from plasmas where chemical potentials are important,
such as stellar cores.

Dark photons provide the simplest example of these effects, which
were treated correctly in some, though not all, of the previous
literature. In Section~\ref{sec:dpsn}, we estimate the production
of dark photons in core-collapse supernovae, previous calculations
of which ignored mixing effects~\cite{Dent:2012mx,Kazanas:2014mca,Rrapaj:2015wgs}, and use this to place new
constraints in the mass-mixing parameter space. Note that there are very probably
extra constraints one could place from arguments beyond
simple energy-loss-from-core~\cite{Kazanas:2014mca}, but we leave those to future work. We also
update the stellar-cooling constraints to include trapping effects, for
heavy and more-strongly-coupled dark photons.

Extending these results to other types of new vectors, we illustrate
the effects in the case of a weakly-coupled $B-L$ vector. In
Section~\ref{sec:blbounds}, we estimate $B-L$ stellar cooling bounds,
correcting previous literature~\cite{Carlson:1986cu,Harnik:2012ni,Heeck:2014zfa}.
Weakly-coupled scalars can have an in-medium mixing with
longitudinal photon oscillations. Resonant production of such
scalars, in particular from the cores of RG stars, places
significantly stronger constraints (Section~\ref{sec:scalarbounds}, Figure~\ref{fig:hp1}) than
previous calculations which ignored mixing effects.
Other BSM particle candidates which could result in similar effects include
vectors coupling to other SM current (though 
the non-conservation of these currents generally leads to strong
	constraints from high-energy experiments),
other forms of scalar coupling, and higher-spin new particles (e.g.\
KK gravitons, as considered in \cite{Hanhart:2001fx}).
We leave the calculation of such constraints to future work.

Since the low-energy SM preserves parity, axion-like particles
do not mix with SM plasma excitations, up to effects suppressed
by weak-scale masses. However, if there is a parity-violating
background, such as a strong magnetic field, then plasma
mixing effects may become important. Nevertheless, even
for a supernova, which may have magnetic fields of order $10^{15}$ Gauss~\cite{Wheeler:2004ff},
rough estimates indicate that this effect will be sub-leading.
The temperature in a SN core is also hot enough that mixing
with pions will have a small effect, but the pion mass is large enough
that this will also be sub-dominant to other processes.

Up until the recombination era (i.e.\ for temperatures
$\gtrsim \eV$), the universe was a hot, basically homogeneous plasma,
so our calculations of the production
rate for light, weakly coupled particles from this plasma will apply.
Such production can also, with some assumptions, be used
to place constraints on the properties of light particles,
and as per stellar cooling calculations, plasma effects
may make an important difference.
For example, the cosmological bounds on a light $B-L$ vector,
as shown in Figure~\ref{fig:bl1}, are from the late-time decay of
a small relic $B-L$ abundance produced in the early universe,
and at the masses indicated, this abundance is dominated by
resonant production around electron-positron freeze-out~\cite{Redondo:2008ec}.
However, the early universe differs from stellar cores
in a number of ways --- chemical potentials were probably very small, and
variation was in time instead of space --- so
the new particle candidates for which plasma effects make an important
difference to observable quantities are not necessarily the same
as those for stellar cooling bounds.
We leave calculations of plasma effects on such early universe
physics to future work.\footnote{There have been a number of papers
investigating the early-universe cosmology of dark photons,
including~\cite{Redondo:2008ec,Fradette:2014sza,Berger:2016vxi}.}

`Medium effects' of the kind discussed in this paper are sometimes
important in laboratory searches for BSM particles, particularly at
very low masses (e.g.~\cite{Dubovsky:2015cca,Chaudhuri:2014dla}).
If we are interested in particle masses well above
the plasma frequency of bulk materials, $\sim \eV$,
then bulk medium effects will be negligible.
For particle production in processes involving nuclei,
the effective `plasma frequency' of the nuclear `medium',
which we expect to be $\sim 10 \MeV$, may be significant.
For example, the beam dump constraints
on dark photons shown in Figure~\ref{fig:sn1} come from the production
of dark photons in high-energy collisions between electrons/protons
and target nuclei.
However, the very small size of nuclei introduces 
important differences with respect to the stellar
production case; in particular, nuclei are optically thin to SM photons
as well as to dark photons. Thus, the previous arguments about only needing
to consider the production of the weakly-coupled in-medium state
will not hold. Further work would be required to calculate
the nuclear medium effects on particle production in experiments such as
beam dumps.

An important point is that the mixing effects giving rise to resonant production
etc are better viewed as `plasma' effects, rather than `thermal' ones.
In particular, although we have used the apparatus of thermal field theory to calculate
them, they will also occur out of thermal equilibrium. In Appendix~\ref{ap:eom},
we show how modified production rates can be derived by considering
classical plasma oscillations in the fluid approximation, including the presence
of a new weakly-coupled field in addition to electromagnetism.
This serves as an independent check of our thermal field theory calculations,
as well as indicating how to extend them to other situations.

Future observations of stars, and of the early universe, will
improve our sensitivity to anomalous energy losses,
and if deviations from Standard Model predictions are found, may
provide hints of new, light BSM particles. In either case,
correct calculations of BSM production rates are vital
in translating such observations into knowledge about new physics.


\section*{Note added}
In the final stages of preparing this manuscript, the
paper~\cite{Chang:2016ntp} which also studies plasma effects on
supernova dark photon bounds, was posted on the arXiv.

\section*{Acknowledgements}

We acknowledge helpful discussions with Asimina Arvanitaki, Masha Baryakhtar,
Anthony Fradette,
Junwu Huang, Maxim
Pospelov, and Yu-Dai Tsai. 
Research at Perimeter Institute is supported by the Government 
of Canada through Industry Canada and by the 
Province of Ontario through the Ministry of Economic 
Development \& Innovation.

\appendix

\section{Appendices}


\subsection{Thermal field theory --- real time formalism}
\label{ap:tft}

In the real-time formalism of thermal field theory~\cite{Bellac:2011kqa},
for each physical field $\phi_1$ we introduce a `ghost' field
$\phi_2$, with the in-medium propagator being a matrix
that mixes the $1,2$ fields. This doubling of degrees
allows the formalism to keep track of the analytic
structure of the contour along which the time integral
is evaluated. We can write the full in-medium propagator $D_{ab}$ 
(with $a,b \in \{1,2\}$) as
\begin{equation}
	D_{ab}^{-1} = (D^F)^{-1}_{ab} + i \Pi_{ab} ~,
\end{equation}
where $D^F$ is the free-field propagator in the thermal bath,
and $\Pi_{ab}$ is the self-energy.
From~\cite{Bellac:2011kqa}, the propagator can be diagonalised 
in the $\{1,2\}$ basis
by the matrix
\begin{equation}
	U = \begin{pmatrix}
		\sqrt{1 + n(\omega)} & \sqrt{n(\omega)} \\
		\sqrt{n(\omega)} & \sqrt{1 + n(\omega)}
	\end{pmatrix} ~,
\end{equation}
giving
\begin{equation}
	D_{ab}(K) = U(\omega)\begin{pmatrix}
		\frac{i}{K^2 - m^2 - \Pi(K)} & 0 \\
		0 & 
		\frac{-i}{K^2 - m^2 - \Pi^*(K)} & 0 \\
	\end{pmatrix}
	U(\omega) ~,
\end{equation}
for a single species of mass $M$, where $\Pi(K)$ is the `real-time'
self-energy.


\subsection{Damping rates in a dilute non-relativistic plasma}
\label{ap:damp}

For a vector coupling to electrons with coupling $g$, the
damping rate from Thomson scattering is~\cite{Redondo:2013lna}, writing $\alpha_g = g^2 / 4 \pi$, 
\begin{equation}
	\Gamma_T \simeq \sigma_L \simeq \frac{8 \pi \alpha \alpha_g n_e}{3 m_e^2}
	\sqrt{1 - \omega_p^2 / \omega^2} ~,
\end{equation}
for $\omega > \omega_p$, and zero for smaller $\omega$ (since
$\omega_p$ is the minimum energy carried by a SM photon in the plasma).

The damping rate from electron-ion bremsstrahlung is
\begin{equation}
	\Gamma_T \simeq \frac{16 \pi^2 \alpha^2 \alpha_g}{3 m_e^2 \omega^3}
	\sqrt{\frac{2 \pi m_e}{3 T}} n_e n_i Z_i^2 \bar{g}_i(\omega,T) \frac{1}{1 + f_B(\omega)} ~,
\end{equation}
to lowest order in $m_e/m_i$, where $\bar{g}$ is the thermally-averaged
Gaunt factor~\cite{Brussaard:1962zz}. In the Born approximation,
\begin{equation}
	\bar{g} = \frac{\sqrt{3}}{\pi} e^{\omega / (2 T)} K_0 (\omega / (2 T)) ~,
\end{equation}
(this approximation is valid for $\omega \ll T$, but $\bar{g}$ 
is order-1 throughout).
In the simplest case of emitting a massless vector, integrating over $\omega$
gives a power per unit volume of
\begin{equation}
	\int \frac{d^3 k}{(2 \pi)^3} f_B(\omega) \omega \Gamma_T
	\simeq \sqrt{\frac{2 \pi}{3}} \frac{8 \alpha^2 \alpha_g n_e n_i Z^2 \sqrt{T}}{3 m_e^{3/2}}
	\langle \bar{g} \rangle ~.
	\label{eq:bremPower}
\end{equation}
The leading-order expressions for $\sigma_L$ are the same.

For a vector coupling to nucleons, we also have ion-ion bremsstrahlung.
The situation here is slightly more complicated, since
the ions interact through a \emph{repulsive} Coulomb potential, rather than
an attractive one as for electron-ion collisions. In the Born approximation,
the damping rates are as for the attractive case, if we substitute
$(X_1/m_1 - X_2/m_2)^2$ for $1/m_e^2$, and the reduced mass
$\mu$ for the rest of the $m_e$, where $m_1$ and $m_2$ are the masses
of the ions, and $X_1,X_2$ their charges under the new vector.

If the ion thermal velocities are low enough --- in particular,
if $Z_1 Z_2 e^2 / v_i \gg 1$, where $v_i \sim \sqrt{T/m_i}$ is a typical
ion thermal velocity --- then the ion wavefunctions are significantly
distorted during the collision, and the Born approximation ceases to hold.
In this regime, the emission can be treated as approximately classical.
\cite{LandauFields} gives the classical dipole radiation spectrum from collisions 
with a given initial approach velocity $v_0$; for high frequencies,
this is suppressed by
\begin{equation}
	\exp\left(-\frac{\pi \alpha Z_1 Z_2}{v_0} \frac{\omega}{\mu v_0^2/2}\right) ~.
\end{equation}
So, in a thermal bath, emission at frequencies $\gg T \frac{v_0}{\alpha Z_1 Z_2}$
is exponentially suppressed --- the Coulomb repulsion between
the ions stops them from getting close enough to undergo
larger accelerations. We can calculate the classical radiative
energy loss analytically, obtaining
\begin{equation}
	\frac{dP}{dV} \simeq \frac{8 \pi}{3} \alpha \alpha_g Z_1 Z_2 \left(\frac{X_1 m_2 -
	X_2 m_1}{\sqrt{m_1 m_2}}\right)^2 \frac{T n_1 n_2}{m_1 m_2} ~,
\end{equation}
for transverse radiation of a massless vector.
If the plasma screening length is comparable to the ion separations
during collisions, $r_0 \sim \alpha Z_1 Z_2 / T$, then screening effects
will become important.


\subsection{Photon self-energies in relativistic/degenerate plasmas}
\label{ap:photonpi}

This Appendix mostly reviews results from~\cite{Braaten:1993jw}, 
both to collect them in a convenient form, and to 
allow us to extend them to mixing self-energies in Appendix~\ref{ap:scalar}.

Evaluating the real part of the one-loop electron self-energy diagram,
using free field propagators for the electrons, we obtain
\begin{align}
	\Pi_r^{\mu\nu} (K) = e^2 \int \frac{d^3 p}{(2 \pi)^3} &\frac{1}{2 E_p}
	(f_e (E_p) + f_{\bar{e}}(E_p)) \nonumber \\
	&\frac{P \cdot K (P^\mu K^\mu + K^\mu P^\nu) - K^2 P^\mu
	P^\nu - (P \cdot K)^2 g^{\mu\nu}}{(P \cdot K)^2 - (K^2)^2 / 4} ~,
	\label{eq:piAA}
\end{align}
which is correct to order $\alpha$.
As discussed in~\cite{Braaten:1993jw, Raffelt:1996wa},
the $K^4$ term in the denominator gives an $\OO(\alpha^2)$ correction
everywhere on the dispersion relation (`on-mass-shell').
It can therefore be ignored at leading
order --- this also prevents the on-mass-shell self-energy
from gaining a $\OO(\alpha^2)$ imaginary part due
$\gamma \rightarrow e^+ e^-$ decays, which
are prevented by the electron also gaining a thermal mass~\cite{Braaten:1993jw}.

Ignoring the $K^4$ term, we can do the angular parts of the integral,
obtaining
\begin{equation}
	\Pi_{T,r} (\omega,k) = 
	\frac{4 \alpha}{\pi}\int_0^\infty dp \, f_p \frac{p^2}{E^2}
	\left(\frac{\omega^2}{k^2} - \frac{\omega^2 - k^2}{k^2} \frac{\omega}{2 k v}
	\log \left(\frac{\omega + v k}{\omega - vk}\right)\right) ~,
\end{equation}
\begin{equation}
	\Pi_{L,r} (\omega,k) = 
	\frac{4 \alpha}{\pi} \frac{\omega^2 - k^2}{k^2} \int_0^\infty dp \, f_p \frac{p^2}{E^2}
	\left(\frac{\omega}{k v}
	\log \left(\frac{\omega + v k}{\omega - vk}\right) - \frac{\omega^2 - k^2}{\omega^2 - k^2 v^2} - 1\right) ~.
\end{equation}
Approximating the energy integral by taking it to be dominated
by a particular electron velocity $v_*$, we obtain the
analytic approximations~\cite{Braaten:1993jw}
\begin{equation}
	\Pi_{T,r}(\omega,k) \simeq \omega_p^2 \frac{3}{2 v_*^2} \left(\frac{\omega^2}{k^2} -
	\frac{\omega^2 - v_*^2 k^2}{k^2} \frac{\omega}{2 v_* k} \log\left(\frac{
		\omega + v_* k}{\omega - v_* k}\right)\right) ~,
\end{equation}
\begin{equation}
	\Pi_{L,r}(\omega,k) \simeq \frac{K^2}{k^2}\omega_p^2 \frac{3}{ v_*^2} \left( \frac{\omega}{2 v_* k} \log\left(\frac{
		\omega + v_* k}{\omega - v_* k}\right) - 1\right) ~,
\end{equation}
where
\begin{equation}
	\omega_p^2 \equiv \frac{4 \alpha}{\pi} \int_0^\infty dp \frac{p^2}{E} f_p \left(1 - \frac{1}{3} v^2\right) ~,
\end{equation}
\begin{equation}
	\omega_1^2 \equiv \frac{4 \alpha}{\pi} \int_0^\infty dp \frac{p^2}{E} f_p 
	\left(\frac{5}{3}v^2 - v^4\right) ~,
\end{equation}
\begin{equation}
	v_* \equiv \omega_1 / \omega_p ~.
\end{equation}
These expression are accurate to order $\alpha$ in the degenerate ($T \ll \mu - m_e$),
`classical' ($m_e - \mu \gg T$),
and relativistic ($T \gg m_e$ or $\mu \gg m_e$) limits.
As per~\cite{Raffelt:1996wa}, if we define the functions
\begin{equation}
	G(x) = \frac{3}{x}\left(1 - \frac{2 x}{3} - \frac{1 - x}{2 \sqrt{x}} 
	\log \left(\frac{1 + \sqrt{x}}{1 - \sqrt{x}}\right)\right) ~,
	\label{eq:gfn}
\end{equation}
\begin{equation}
	H(x) = \frac{G(x) - x}{x - 1}
	=   \frac{3}{2 x^{3/2}}
	\log \left(\frac{1 + \sqrt{x}}{1 - \sqrt{x}}\right)
- 1 - \frac{3}{x} ~,
	\label{eq:hfn}
\end{equation}
then the above expressions are
\begin{equation}
	\Pi_T \simeq \omega_p^2 \left(1 + \frac{1}{2} G(v_*^2 k^2 / \omega^2)\right) ~,
\end{equation}
\begin{equation}
	\Pi_L \simeq \frac{K^2}{\omega^2} \omega_p^2 (1 +  H(v_*^2 k^2 / \omega^2)) ~.
 \label{eq:pil}
\end{equation}
We see that the dispersion relations both have $\omega = \omega_p$ at $k = 0$,
and that the frequency at which the longitudinal dispersion 
relation crosses the lightcone goes to infinity as the electron velocity
goes to one. However, the crossing frequency increases rather slowly.
For example, in a SN core, with $T \simeq 30 \MeV$ and $\mu_e \simeq 350 \MeV$,
we have $v_* \simeq 1 - 2 \times 10^{-6}$. On the light-cone,
this gives $1 + H(v_s^2) \simeq 18.7$, and since
$\omega_p \simeq 20 \MeV$, the crossing frequency is $\sqrt{(1 + H(v_*^2))} \omega_p
\simeq 85 \MeV$. In general,
\begin{equation}
	H(1 - \delta^2) \simeq 3 \log \delta^{-1} + \dots
\end{equation}
 
\begin{figure}
	\begin{center}
		\includegraphics[width=.48\textwidth]{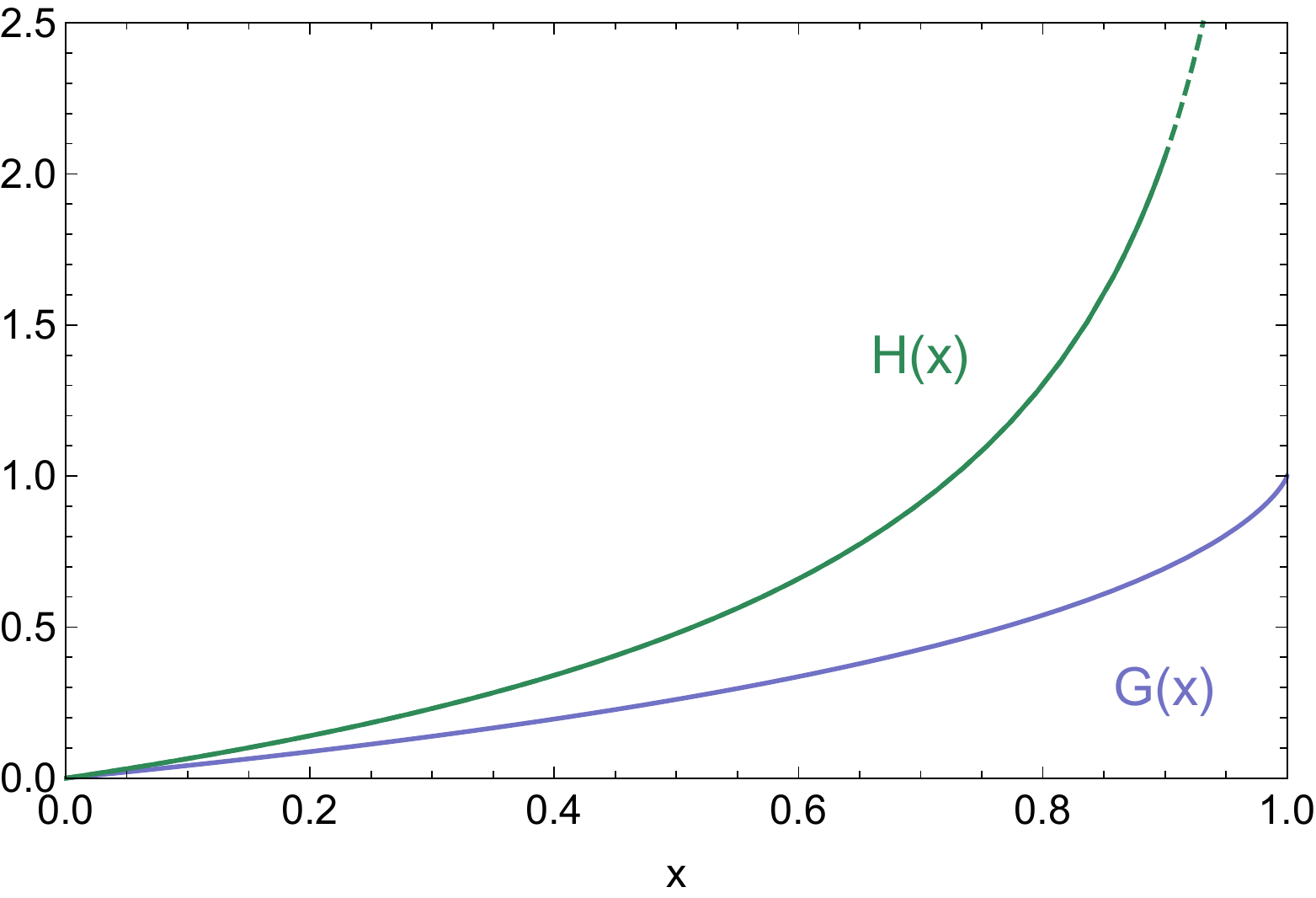}
	\end{center}
	\caption{$G(x)$ (blue curve) and $H(x)$ (green curve), as defined
	in equation~\ref{eq:gfn} and~\ref{eq:hfn}.}
	\label{fig:gfn}
\end{figure}

To evaluate resonant longitudinal production
rates (e.g.~\ref{eq:resL}), we need the expression
\begin{equation}
	1 - \left.\frac{d\omega_L^2}{d\omega^2}\right|_{K^2 = m^2} \simeq
	1 + \frac{m^2}{2 k^2} \left(3 + \frac{2 x^{3/2}}{(1-x)\left(2 \sqrt{x} - 
	\log \left(\frac{1 + \sqrt{x}}{1 - \sqrt{x}}\right)\right)
	}\right) ~,
\end{equation}
where $x = v_*^2 k^2 / \omega^2$.
For $x = 1 - \delta^2$, this is 
\begin{equation}
	\simeq 1 - \frac{m^2}{2 k^2 \delta^2 \log \delta^{-1}} + \dots 
\end{equation}
Since $\delta > m^2/\omega^2$, this factor is always $\OO(1)$.


\subsection{Scalar self-energies}
\label{ap:scalar}

The one-loop electron self-energy expressions for $\phi \phi$ and
$\phi$-photon mixing, for a $g_\phi \phi \bar{e} e$ coupling, are
\begin{align}
	(\Pi^\mu)^{\phi A} (K) = g_\phi e \int \frac{d^3 p}{(2 \pi)^3} &\frac{1}{2 E_p}
	(f_e (E_p) + f_{\bar{e}}(E_p)) \nonumber \\
	& m_e \frac{(P \cdot K) K^\mu - K^2 P^\mu}{(P \cdot K)^2 - (K^2)^2 / 4} ~,
	\label{eq:phiAL}
\end{align}
\begin{align}
	\Pi^{\phi\phi} (K) = g_\phi^2 \int \frac{d^3 p}{(2 \pi)^3} &\frac{1}{2 E_p}
	(f_e (E_p) + f_{\bar{e}}(E_p)) \nonumber \\
	& \frac{(P \cdot K)^2 - m_e^2 K^2}{(P \cdot K)^2 - (K^2)^2/4} ~,
\end{align}
(as compared to the photon-photon expression in equation~\ref{eq:piAA}).
By the same arguments as in Appendix~\ref{ap:photonpi}, if we are just
interested in the behaviour on the photon mass shell, or for $K^2$
sufficiently small, we can ignore the $K^4$ term in the denominator,
giving
\begin{equation}
	\Pi^{\phi L} \simeq \frac{g e m_e}{2 \pi^2 k} \sqrt{K^2} \int_0^\infty dp \, v^2 f_p \left(
	\frac{\omega}{v k} \log \left(\frac{\omega + v k}{\omega - v k}\right)
	- 2 \frac{K^2}{\omega^2 - k^2 v^2}\right) ~.
\end{equation}
where we have written $\Pi^{\phi L} \equiv \epsilon_L^\mu (\Pi_\mu)^{\phi A}$.
For a non-relativistic plasma, this gives
\begin{equation}
	\Pi^{\phi L} \simeq g e \frac{k \sqrt{K^2}}{\omega^2} \frac{n_e}{m_e} ~,
\end{equation}
in agreement with the expression in Section~\ref{sec:scalarprod}.
The $\Pi^{\phi L}_r$ self-energy is related to the resonant scalar production
rate by
\begin{equation}
	\frac{d\dot N}{dV} \simeq \frac{1}{4\pi} \left(\frac{\omega_L}{m} \Pi^{\phi L}_r\right)^2
	\frac{1}{e^{\omega_L/T} - 1} 
			\left| 1- \left. \frac{d\omega_L^2}{d\omega^2}\right|_{\omega_L}\right|^{-1} ~.
\end{equation}


\subsection{Decay to electrons and positrons}
\label{ap:ep}

In Appendix~\ref{ap:photonpi}, we noted that for a photon excitation of
small invariant mass, the in-medium mass of the electron prevents the
decay $\gamma \rightarrow e^+ e^-$. However, if the new
weakly coupled state has mass $m > 2 \tilde{m}_e$, where $\tilde{m}_e$ is the effective electron
mass, the weakly-coupled state can decay into an electron and a positron,
and we need the corresponding imaginary parts
of $\Pi^{XX}$, $\Pi^{XA}$ and $\Pi^{AA}$ to evaluate the absorption
rate from this decay.
The simplest case, and the only one which is relevant to the calculations
in this paper, is that of a heavy dark photon, where we only need to calculate
$\Pi^{AA}_i$ at large invariant mass.

From~\cite{Braaten:1991hg}, the effective electron mass (at $k=0$) is
\begin{equation}
	\tilde{m}_e = \frac{m_e}{2} + \sqrt{\frac{m_e^2}{4} + m_{\rm eff}^2} ~,
\end{equation}
where
\begin{align}
	m_{\rm eff}^2 &= \alpha \int_0^\infty E dE (f_-(E) + f_+(e) + 2 f_B(E))
	\\
	&= \alpha \int_0^\infty E dE \left(
	\frac{1}{e^{(E-\mu)/T} + 1}
	+ \frac{1}{e^{(E+\mu)/T} + 1}
	+ 2 \frac{1}{e^{E/T} - 1}
	\right) ~.
\end{align}
For example, in a SN core with $T = 30 \MeV$, $\mu = 350 \MeV$, we have 
$\tilde{m}_e \simeq 12 \MeV$.
For an ultra-relativistic electron, the dispersion relation
has $\omega^2 - k^2$ ranging from $m_{\rm eff}^2$ at $k=0$
to $2 m_{\rm eff}^2$ as $k\rightarrow \infty$.
The decay rate for longitudinal excitations is (ignoring the
longitudinal residue factor, and in the rest frame of the plasma)
\begin{equation}
	\Gamma_{L, {\rm dec}}(\omega,k) \simeq  \frac{\alpha}{2 \omega k} K^2
	\int_{E_-}^{E^+} dE \, g(E)
	\left(1 - 4 \left(\frac{E}{k} - \frac{\omega}{2 k}\right)^2\right)
	(1 - f_-(E)) ~,
\end{equation}
where $f_-(E) = 1/(e^{(E -\mu)/T}+1)$ is the electron occupation number, giving
Pauli blocking, $g(E)$ is an order-1 function summarising the in-medium residue factor
for the electron/positron, and $E_\pm$ are the maximum and minimum decay energies
in the plasma rest frame. Ignoring the frequency variation in the
electron effective mass, $E_\pm = (1/2)\left(\omega \pm \beta_{cm} k\right)$,
where $\beta_{cm} \equiv \sqrt{1 - 4 \tilde{m}_e^2 / m^2}$.
For transverse excitations (again ignoring the transverse residue factor),
\begin{equation}
	\Gamma_{T, {\rm dec}}(\omega,k) \simeq  \frac{\alpha}{2 \omega k} K^2
	\int_{E_-}^{E^+} dE \, g(E)
	\left(1 + 2 \left(\frac{E}{k} - \frac{\omega}{2 k}\right)^2
	- \frac{\beta_{cm}^2}{2}\right)
	(1 - f_-(E)) ~.
\end{equation}
Ignoring Pauli-blocking (i.e.\ setting $f_- = 0$) and
residue factors,
we obtain
\begin{equation}
	\Gamma_{T,{\rm dec}} = \Gamma_{L,{\rm dec}} =
	\frac{\alpha}{3} \frac{m^2}{\omega} \beta \left(1 + \frac{2 m_e^2}{m^2}\right) ~,
\end{equation}
which is the correct decay rate in vacuum.

\subsection{Dark photon production and absorption in supernovae}
\label{ap:reddy}

Here we give cross sections for dark photon production and absorption
in the soft approximation,
following \cite{Rrapaj:2015wgs} but adapting to include
thermal mixing effects (which requires separating the longitudinal and
transverse cases).

Assuming the energy of an emitted photon is small compared to the energy
of the proton-neutron interaction, the cross section for the emission of
a photon can be related to that for elastic neutron-proton scattering by
\begin{equation}
d\sigma_{pn\rightarrow pn\gamma} \simeq -4 \pi \alpha \left(\epsilon_{\mu} J^{\mu}\right)^2 \frac{d^3k}{2\omega} d\sigma_{pn\rightarrow pn} ,
\end{equation}
where $J$ is the dipole current
\begin{equation}
J = \frac{P_1}{P_1.K} - \frac{P_2}{P_2.K} ~,
\end{equation}
with $K$ the four--momentum of the emitted photon, and $P_1$ and
$P_2$ the initial and final four--momentum of the proton. 
As well as a dipole contribution from neutron--proton
interactions there will also be quadrupole pieces from proton--proton
interactions, however these are smaller.

For the longitudinal polarisation we have 
\begin{equation}
\frac{1}{4\pi} \int d\Omega_k \left(\epsilon\cdot J\right)^2 \simeq \frac{2}{3}\frac{E_{cm}}{m_n} \frac{K^2}{\omega^4} \left(1- \cos \theta_{cm} \right) .\end{equation}
in the centre of mass frame, and assuming the nucleons are
non-relativistic. This is suppressed by $m^2/\omega^2$ and in
the limit $m\ll \omega \sim T$ is small as expected. For the transverse modes we have, summing
over the two polarisations,
\begin{equation}
\frac{1}{4\pi} \int d\Omega_k \sum_i \left(\epsilon_i \cdot  J\right)^2 \simeq \frac{4}{3}\frac{E_{cm}}{m_n} \frac{1}{\omega^2} \left(1- \cos \theta_{cm} \right) .
\end{equation}
Assuming nucleons are non-degenerate in the supernova environment, the
initial proton and neutron states can be integrated over. The rate of energy loss per unit volume, by continuum emission of dark
photons is then given by
\begin{equation}
\frac{dE}{dVdt} = \frac{4 \alpha n_n n_p}{\left(\pi m_n T\right)^{3/2}} \int_{m}^{\infty} dE  ~ E^3 e^{-E/T}  I_{\left(L,T\right)}\left(E\right) \sigma_{np}\left(E\right) ~,
\end{equation}
where $m$ is the dark photon mass. The quantity 
\begin{equation}
\sigma_{np}\left(E\right) = \int d \cos\theta_{cm}~\left(1-\cos\theta_{cm}\right) \frac{d\sigma_{np}}{d\cos\theta_{cm}} ~,
\end{equation}
can be extracted from data and is given in \cite{Rrapaj:2015wgs}. The
functions $I_{L,T}$ include the effects of mixing and are given by
\begin{equation}
I_T\left(E\right) = \epsilon^2 \frac{m^4}{\left(m_T^2-m^2\right)^2 } \frac{4}{3} \left(\sqrt{1- \left(\frac{m}{E}\right)^2} - \left(\frac{m}{E}\right) \arccos\left(\frac{m}{E}\right) \right)  ~,
\end{equation}
and
\begin{equation}
I_L\left(E\right) = \epsilon^2 \int_m^E d\omega~  \frac{\omega^4}{\left(\omega_L^2-\omega^2\right)^2 } \frac{2 \sqrt{\omega^2-m^2} m^2}{3 \omega^3 E} ~,
\end{equation}
where the notation is the same as in Section~\ref{sec:smplasma}, with
$\omega_L$ given by equation~\eqref{eq:pil}. As discussed in Section~\ref{sec:dp}, for resonant
emission the value of the imaginary part of the photon self energy cancels out when computing
the overall energy emitted.

The absorption cross sections are obtained analogously, and for the dark
photon the inverse mean free path is
\begin{equation}
\frac{1}{\lambda_T} = \frac{32}{3\pi} \alpha n_n n_p \left(\frac{\pi T}{m_n} \right)^{3/2} \frac{1}{\omega^3 \sqrt{1-\left(\frac{m}{\omega}\right)^2}}  \frac{\epsilon^2 m^4}{\left(m_T^2-m^2\right)^2 } \left<\sigma_{np}\left(T\right)\right> ~,
\end{equation}
for transverse modes, and
\begin{equation}
\frac{1}{\lambda_L} =  \frac{32}{3\pi} \alpha n_n n_p \left(\frac{\pi T}{m_n} \right)^{3/2} \frac{m^2}{\omega \sqrt{1-\left(\frac{m}{\omega}\right)^2}}   \frac{\epsilon^2}{\left(\omega_L^2-\omega^2\right)^2 } \left<\sigma_{np}\left(T\right)\right> ~,
\end{equation}
for longitudinal modes. Here 
\begin{equation}
\left<\sigma_{np}\left(T\right)\right> = \int_0^\infty dx \frac{1}{2} x^2 e^{-x} \sigma_{np}\left(x T\right) ~, 
\end{equation}
is a thermally
averaged cross section, and numerical values as a function of
temperature are again given in \cite{Rrapaj:2015wgs}.\footnote{Our expressions match those of \cite{Chang:2016ntp} and differ from those of \cite{Rrapaj:2015wgs} by numerical factors.}


\subsection{Classical equations of motion}
\label{ap:eom}

To illustrate how our `mixing effects' arise outside
the thermal field theory formalism, we can look at the classical
EoM for plasma oscillations in the fluid approximation, with the presence of an additional weakly-coupled
field (as per e.g.~\cite{Dubovsky:2015cca}).
For example, considering the case of longitudinal oscillations in the presence
of a $\phi \bar{e} e$ scalar, the electric field $E$ and scalar field $\phi$
are sourced by (assuming that ions have $Z = 1$)
\begin{equation}
	\vec{\nabla} \cdot \vec{E}  = - e (n_e - n_i)
	\quad , \quad
	(\partial_t^2 - \nabla^2 - m^2) \phi \simeq g n_e ~,
\end{equation}
where we have assumed that the electron motions
are non-relativistic. The continuity equations are
\begin{equation}
	\dot n_e + \vec{\nabla} \cdot (n_e \vec{v}_e) = 0
	\quad , \quad
	\dot n_i + \vec{\nabla} \cdot (n_i \vec{v}_i) = 0 ~,
\end{equation}
and the momentum equations (ignoring thermal diffusion) are
\begin{equation}
	m_e n_e (\dot \vec{v}_e + \vec{v}_e \cdot \vec{\nabla} \vec{v}_e) \simeq - n_e (e \vec{E} - g \vec{\nabla} \phi) - m_e n_e \nu (\vec{v}_e - \vec{v}_i) ~,
\end{equation}
\begin{equation}
	m_i n_i (\dot \vec{v}_i + \vec{v}_i \cdot \vec{\nabla} \vec{v}_i) \simeq  n_i e \vec{E} + m_e n_e \nu (\vec{v}_i - \vec{v}_e) ~,
\end{equation}
where we have included a `frictional drag' term from electron-ion
collisions,
parameterised by an `effective collision frequency' $\nu$
(we have ignored magnetic fields, since the velocities are non-relativistic).

To find the equation of motion for small oscillations, we linearise
about the equilibrium values of quantities, writing
$n_e = n_0 (1 + \delta_e)$,
$n_i = n_0 (1 + \delta_i)$, 
$\phi = \phi_0 + \delta\phi$, and 
and $v_e,v_i,E$ as small quantities themselves.
Taking the limit where the ion mass is much heavier than the electron
mass, we can set $v_i, \delta_i = 0$ throughout. 
Writing the linearised quantities as $e^{- i (\omega t -  k x)}$ (i.e. taking propagation
to be in the $x$ direction), and solving for $v_e, \delta_e$ in terms of $E, \delta\phi$,
we obtain the equations
\begin{equation}
	\left(
	\begin{pmatrix}
		-i k & 0 \\
		0 & \omega^2 - k^2 - m^2 
	\end{pmatrix}
	- \frac{n_0}{m_e} \frac{k}{\omega (\omega - i \nu)}
	\begin{pmatrix}
		-i e^2 & e g k \\
		i e g & - g^2 k
	\end{pmatrix}
	\right)
	\begin{pmatrix}
		E \\ \delta\phi
	\end{pmatrix}
	= 0 ~.
\end{equation}
For $g = 0$, we obtain the separate equations
\begin{equation}
  1 - \frac{\omega_p^2}{\omega(\omega - i \nu)} 
	\quad \Rightarrow \quad 
	\omega = \sqrt{\omega_p^2 - \frac{\nu^2}{4\omega_p^2}} + \frac{i \nu}{2} ~,
\end{equation}
for $E$, and $\omega^2 - k^2 - m^2 = 0$ for $\delta\phi$, as expected.
With $g \neq 0$, if we write $\omega = \sqrt{k^2 + m^2} + g^2 \omega_g \equiv \omega_0 + g^2 \omega_g$, then
the weakly-coupled mode has
\begin{equation}
	\Imag \, \omega_g = - k^2 \frac{n_0}{2 m_e} \frac{\nu}{\omega_0^2 \nu^2 + (\omega_0^2 - \omega_p^2)^2} ~.
	\label{eq:eomim}
\end{equation}
Around the resonance, when $\omega_0 \simeq \omega_p$, this has the form
of equation~\ref{eq:phieenr}, since as per above, $\sigma_L \simeq \nu$
there. This confirms that the resonant production
is not $m$-suppressed for small $m$. 
Inserting the general relationship between
$\nu$ and $\sigma_L$ into equation~\ref{eq:eomim}, we obtain equation~\ref{eq:phieenr},
as required.

Analogous calculations can be done for the new particle candidates discussed
in this work, replicating the results of the thermal field theory calculations
as required.


\bibliography{thermalProduction}
\bibliographystyle{JHEP}

\end{document}